\documentclass[entropy,article,accept,pdftex,moreauthors]{Definitions/mdpi} 

\usepackage{amssymb}
 \usepackage{amsthm}
 \usepackage{amsmath}
\usepackage{mathrsfs}
\usepackage{bbm}

\usepackage{graphicx}
\usepackage{dcolumn}
\usepackage{bm}
\usepackage{amsfonts}    
\usepackage{tikz-cd}
\usepackage{comment}

\firstpage{1} 
\pubvolume{1}
\issuenum{1}
\articlenumber{0}
\pubyear{2023}
\copyrightyear{2023}
\datereceived{ } 
\daterevised{ } 
\dateaccepted{ } 
\datepublished{ } 
\hreflink{https://doi.org/} 

\Title{The Thermomajorization Polytope and Its Degeneracies}

\TitleCitation{The Thermomajorization Polytope and Its Degeneracies}

\Author{Frederik vom Ende $^{1,\dagger}$\orcidA{} and
Emanuel Malvetti $^{2,3,\dagger}$\orcidB{}}

\AuthorNames{Frederik vom Ende and Emanuel Malvetti}

\AuthorCitation{vom Ende, F.; Malvetti, E.}

\address{%
$^{1}$ \quad Dahlem Center for Complex Quantum Systems, Freie Universit{\"a}t Berlin, Berlin, 14195, Germany; frederik.vom.ende@fu-berlin.de\\
$^{2}$ \quad School of Natural Sciences, Technische Universit{\"a}t M{\"u}nchen, Garching, 85737, Germany; emanuel.malvetti@tum.de\\
$^{3}$ \quad Munich Center for Quantum Science and Technology (MCQST) \& Munich Quantum Valley (MQV), M{\"u}nchen, 80799, Germany}

\firstnote{These authors contributed equally to this work.}

\abstract{
Drawing inspiration from transportation theory, in this work we introduce the notions of ``well-structured'' and ``stable'' Gibbs states and we investigate their
implications for quantum thermodynamics and its resource theory approach via thermal operations.
It turns out that, in the quasi-classical realm, global cyclic state transfers are impossible if and only if the Gibbs state is stable.
Moreover, using a geometric approach by studying the so-called thermomajorization polytope
we prove that any subsystem in equilibrium
can be brought out of equilibrium via thermal operations.
Interestingly, the case of some subsystem being in equilibrium
can be witnessed via degenerate extreme points of the thermomajorization polytope, assuming the Gibbs state of the system is well structured.
These physical considerations are complemented by simple new constructions for the polytope's extreme points as well as for an important class of extremal Gibbs-stochastic matrices. 
}


\begin{document}

%
%

\section{Introduction}
The core idea of quantum thermodynamics---a field which has gained significant traction within the last decade---is to apply thermodynamic ideas to (ensembles of) systems of finite size instead of ``thermodynamically large'' setups \cite{VS16}. 
While this discipline comes with a number of principal questions and concepts (such as fluctuation theorems, thermal machines, the fundamental laws, thermalization, and many more, cf.~\cite{QThermo2018})
an all-round successful approach has been to take the open systems perspective \cite{Kosloff13}:
There one models changes to a system of interest as the reduced action of a larger, closed system (i.e.~system plus some form of environment, such as a bath)
through some total Hamiltonian\footnote{
To avoid any confusion let us point out explicitly that the individual Hamiltonians $H_S,H_B$ are to be understood as local terms $H_S\otimes{\bf1}_E,{\bf1}_S\otimes H_B$, in particular when adding them.
} \mbox{$H_{\rm tot}=H_S+H_B+H_I$}
consisting of a system's, an environment, and an interaction term.
Typical thermodynamic constraints imposed on top are that the environment (bath) starts out in equilibrium, i.e.~in a Gibbs state,
or that the total energy is conserved\footnote{
Be aware that some works have weakened to energy conservation to average energy conservation in the past. This leads to a framework where the quantum free energy characterizes all possible state changes \cite{Skrzypczyk14}.
},
i.e.~$[H_{\rm tot},H_S+H_B]=[H_I,H_S+H_B]=0$.

This perspective ties into the information-theoretic and more specifically into the resource theory approach to quantum thermodynamics.
There one attempts to formalize
which operations can be carried out at no cost (with respect to some resource, e.g., work), and one of the main aspects of this approach is to characterize when state transfers under thermodynamic constraints are possible.
This comes with a subset of quantum channels called \textit{thermal operations}, which can be carried out without having to expend any resource.
They arise from the above open systems construction as follows  \cite{Lostaglio19,Janzing00,vomEnde22thermal}:
Given an $n$-level system described by some system's Hamiltonian
$H_S\in{i}\mathfrak u(n)$ as well as some fixed background temperature $T>0$ which every bath needs to adopt, the set $\mathsf{TO}(H_S,T)$ of all thermal operations with respect to $H_S,T$ is defined as
\begin{align*}
\Big\{
\operatorname{tr}_B\Big(e^{iH_{\rm tot}}\Big((\cdot)\otimes
\frac{e^{-H_B/T}}{\operatorname{tr}(e^{-H_B/T})}\Big)e^{-iH_{\rm tot}}\Big) :\ 
\substack{m\in\mathbb N, H_B\in{i}\mathfrak{u}(m),\,H_{\rm tot}\in i\mathfrak{u}(mn)\\
[H_{\rm tot},H_S+H_B]=0 }\Big\}\,.
\end{align*}
Here $\mathfrak{u}(m)$ is the unitary Lie algebra in $m$ dimensions so ${i}\mathfrak u(m)$ is the collection of Hermitian $m\times m$ matrices.
As hinted at before,
one of the central questions of this framework is the following: Given 
an initial and a target state
of some quantum system,
 can the former be transformed into the 
 latter by means of a thermal operation? 

The resource theory approach to quantum thermodynamics lead to
a number of structural insights, ranging from optimal protocols for work extraction \cite{Skrzypczyk14} and cooling \cite{Alhambra19}
to the so-called second laws of quantum thermodynamics
\cite{Horodecki13,Brandao15}.
%
%
The latter precisely relates to the above state interconversion problem via
a generalization of classical majorization called ``thermomajorization''.
First described by Ruch et al.~\cite{Ruch78}~in 
the 70s, thermomajorization has gained widespread popularity over the last decade 
due to the influential works of Brand\~ao et al.~\cite{Brandao15}, Horodecki \& 
Oppenheim \cite{Horodecki13}, Renes \cite{Renes14}, as well as many
more \cite{Faist17,Gour15,Lostaglio18,Sagawa19,Mazurek19,Alhambra19};
there, among other things, it has been used to solve the state interconversion problem in the quasi-classical realm, more on this down below.
%
%
Yet, one can tackle this problem from another, more geometric perspective:
for this one abstractly defines the collection of all states which can (approximately) be generated via thermal operations
 starting from some initial state $\rho$
 $$
 M_{H_S,T}(\rho):=\big\{ \Phi(\rho) : \Phi\in\overline{\mathsf{TO}(H_S,T)} \big\}\,,
 $$
and then one studies the geometric properties of $M_{H_S,T}(\rho)$.
This set has been called 
\textit{(future) thermal cone}\footnote{
Be aware that the term ``cone'' originates from light cones in general relativity and should not be confused with (convex) cones from linear algebra.
}
\cite{Lostaglio15_2,Korzekwa17,Oliveira22} or, in 
the case of quasi-classical states $\rho$, \textit{thermal polytope} 
\cite{Alhambra19} before, and the elements of $M_{H_S,T}(\rho)$ are 
precisely those states which are said to be \textit{thermomajorized} by $\rho$.
%
%
%
In the quasi-classical case
$[\rho,H_S]=0$
the structure of the thermal cone is known to simplify considerably. 
This is due to the following crucial facts:
\begin{itemize}
\item Every thermal operation leaves the set of quasi-classical states
invariant \cite{Lostaglio15}: if $[\rho,H_S]=0$, 
then $[\Phi(\rho),H_S]=0$ for all $ \Phi\in\overline{\mathsf{TO}(H_S,T)}$.
\item Thermal operations and general Gibbs-preserving quantum channels 
(approximately) coincide on quasi-classical states when considering only the 
diagonal \cite[Sec.~3.4]{Shiraishi20}.
\end{itemize}
Combining these facts shows that 
for any state $\rho$ with $[\rho,H_S]=0$ 
where $H_S$ is non-degenerate
(i.e.~$\rho$ is diagonal in ``the'' eigenbasis of $H_S$),
$M_{H_S,T}(\rho)$ equals the set of all diagonal states $\Phi(\rho)$ where $\Phi$ is a quantum channel that preserves the Gibbs state $e^{-H_S/T}/Z$.
But for diagonal states---where we write $x,y\in\mathbb R^n$ for the vectors of their respective diagonal entries---the existence of such a channel
is equivalent to 
the existence of a 
Gibbs-stochastic matrix\footnote{
Recall that given some Hamiltonian $H_S=\operatorname{diag}(E_1,\ldots,E_n)$ as well as $T>0$ a matrix 
$A$ is called \textit{Gibbs-stochastic} if the entries of $A$ are non-negative, every column of $A$ sums up to one, and $Ad=d$.
Here $d:=(e^{-E_j/T})_{j=1}^n$ is the (unnormalized, but can also be normalized) vector of Gibbs weights.
Moreover, we write $s_d(n)$ for the collection of all Gibbs-stochastic matrices w.r.t.~the Gibbs vector $d$.\label{footnote_gibbs_stoch_mat}
}
$A$ such that $Ay=x$ \cite[Coro.~4.5]{vomEnde20Dmaj}.
Thus
$
M_{H_S,T}(\operatorname{diag}(y))$
equals
\mbox{$\{ \operatorname{diag}(Ay): 
A\in\mathbb R^{n\times n}\text{ Gibbs-stochastic}\}
$}
where one, instead and equivalently, can focus solely on the diagonal by considering the so-called \textit{thermomajorization polytope}
\begin{equation}\label{eq:def_M_d_y}
M_d(y):=\big\{Ay:A\in\mathbb R^{n\times n}\text{ Gibbs-stochastic}\,\big\}\,.
\end{equation}
At first glance focusing on the quasi-classical case may appear fruitless for
understanding the behavior of quantum systems.
However, we point out that 
optimal cooling protocols rely on two level Gibbs-stochastic matrices \cite{Alhambra19} which can be realized within the Jaynes-Cummings model \cite{Lostaglio18}.
Moreover, taking the quasi-classical perspective allows one to identify thermal operations which are simple to implement experimentally \cite{Perry18},
and tools from the quasi-classical case have been used in quantum control theory to find non-trivial upper bounds on reachable states for dissipative bath-couplings \cite{CDC19,OSID_thermal_res}.

In this work we will combine this geometric approach to quantum thermodynamics 
with the established field of transportation theory. 
While this is not the first time these fields are being linked---cf.~Sec.~\ref{sec_transp_pol} for a short review and Sec.~\ref{sec_degen} for some results this perspective has lead to already---this paper's main idea is not to take tools, but rather (not as obvious, yet key) concepts from transportation theory and to investigate their
implications in quantum thermodynamics.
Thus, this work is structured as follows:
We begin by recapping characterizations of thermomajorization in the quasi-classical realm in Sec.~\ref{sec_recap},
followed by a review of the basics of transportation theory
in Sec.~\ref{sec_transp_pol}.
Therein we also introduce (or rather, translate) the key concepts 
of ``stable'' and ``well-structured'' Gibbs states,
which will turn out to be quite intuitive.
The implications of these notions will be the topic of Sec.~\ref{sec_results} which contains the main results.
More precisely, stable Gibbs states turn out to be in one-to-one correspondence to the impossibility of (global) cyclic state transfers---which will also lead us to the notion of a ``sub{space} in equilibrium''---cf.~Sec.~\ref{sec_recap_2}.
The latter notion turns out to be closely connected to the geometry of $M_d(y)$ and is reflected in the (number of) extreme points of $M_d(y)$, assuming the Gibbs state is well structured (Sec.~\ref{sec_degen}).
The point-of-view taken in this work also leads to simple, intuitive ways to construct extreme points as well as Gibbs-stochastic matrices which realize the corresponding state transfer;
this is what the example and visualization Sec.~\ref{sec_ex} will be about.

\section{Preliminaries \& Recap}
\subsection{Thermomajorization for Quasi-Classical States}
\label{sec_recap}

The object related to thermomajorization most 
commonly found in the literature is the following:
let any $d\in\mathbb R_{++}^n$ (that is, $d\in\mathbb R^n$ with $d>0$ as $d$ 
plays the role of the vector of Gibbs weights) as well as $y\in\mathbb R^n$ be 
given. One defines the \textit{thermomajorization curve of $y$ (with respect to 
$d$)}, denoted by $\mathrm{th}_{d,y}$, as the piecewise linear, continuous curve 
fully characterized by the elbow points $\{ (\sum_{i=1}^j d_{\tau(i)},
\sum_{i=1}^j y_{\tau(i)}) \}_{j=0}^n\,$ where $\tau\in S_n$ is any 
permutation such that $\frac{y_{\tau(1)}}{d_{\tau(1)}}
\geq\ldots\geq\frac{y_{\tau(n)}}{d_{\tau(n)}}$.
Equivalently \cite[Remark~7]{vomEnde22}, $\mathrm{th}_{d,y}:[0,
{\bf e}^{\top}d]\to \mathbb R$ for all $c\in [0,{\bf e}^{\top}d]$ satisfies
\begin{equation}\label{eq:thermo_curve_char}
\begin{split}
\mathrm{th}_{d,y}(c)&= \min_{\{i=1,\ldots,n\}} \Big(\Big(
\sum_{j=1}^n\max\Big\{y_j-\frac{y_i}{d_i}d_j,0\Big\}\Big)+\frac{y_i}{d_i}c\Big)\\
&=\frac{{\bf e}^\top y}2  +\min_{\{i=1,\ldots,n\}} \Big(\Big\|y-
\frac{y_i}{d_i}d\Big\|_1 +\frac{y_i}{d_i}\Big(c-
\frac{{\bf e}^\top d}2\Big)\Big)
\end{split}
\end{equation}
where, here and henceforth,
${\bf e}:=(1,\ldots,1)^\top$ and $\|\cdot\|_1$ is the usual vector $1$-norm.
Note that this curve is a generalization of the notion of Lorenz curves from 
majorization theory \cite[p.~5]{MarshallOlkin}.

\begin{Remark}\label{rem_perm_matrix}
It is clear from the definition that the thermomajorization curve is invariant under 
permutations
%
in the sense that $\mathrm{th}_{\underline{\sigma} d,\underline{\sigma} y}
\equiv\mathrm{th}_{d,y}$ for all $\sigma\in S_n$.
Here, given some permutation ${\sigma}\in S_n$ we write $\underline{\sigma}$ for the corresponding 
permutation matrix
$\sum_{i=1}^n e_i e_{{\sigma}(i)}^\top $.
In particular
the identities 
$\underline{{\sigma}\circ\tau}=\underline{\tau}\cdot\underline{{\sigma}}$,
$(\underline{{\sigma}}x)_j=x_{{\sigma}(j)}$, 
and $(\underline{\sigma}A\underline{\tau})_{jk}=A_{\sigma(j)\tau^{-1}(k)}$
hold for all
$A\in\mathbb R^{n\times n}$,
$x\in\mathbb R^n$, $j,k=1,\ldots,n$, and all $\sigma,\tau\in S_n$.
\end{Remark}
Now the precise connection between thermomajorization and (quasi-classical) state transfers is summarized in the following well-known result: Given any $x,y\in\mathbb R^n$ the following statements are equivalent
\cite[Prop.~1]{vomEnde22}.
\begin{itemize}
\item There exists a Gibbs-stochastic matrix $A$ (recall footnote~\ref{footnote_gibbs_stoch_mat}) such that $Ay=x$. We denote 
this by $x\prec_d y$.
\item 
${\bf e}^{\top} x={\bf e}^{\top} y$ and $\mathrm{th}_{d,x}(c)
\leq\mathrm{th}_{d,y}(c)$ for all $c\in[0,{\bf e}^\top d]$.
\item  ${\bf e}^{\top}x={\bf e}^{\top}y$ and
$\mathrm{th}_{d,x}(\sum_{i=1}^j d_{\tau(i)})\leq\mathrm{th}_{d,y}(\sum_{i=1}^j 
d_{\tau(i)})$, that is,
$$
\sum_{i=1}^jx_{\tau(i)}\leq \min_{\{i=1,\ldots,n\}}  \Big(\Big(\sum_{j=1}^n\max
\Big\{y_j-\frac{y_i}{d_i}d_j,0\Big\}\Big)+\frac{y_i}{d_i}\Big(
\sum_{k=1}^jd_{\tau(k)}\Big)\Big)
$$
for all $j=1,\ldots,n-1 $
where $\tau\in S_n$ is any permutation such that
$\frac{x_{\tau(1)}}{d_{\tau(1)}}\geq\ldots\geq
\frac{x_{\tau(n)}}{d_{\tau(n)}}$.
\item $\|x-td\|_1\leq\|y-td\|_1$ for all $t\in\mathbb R$.
\item ${\bf e}^{\top} x={\bf e}^{\top} y$ and $\|d_ix-y_id\|_1\leq\|d_iy-
y_id\|_1$ for all $i=1,\ldots,n $.
\end{itemize}
These criteria slightly simplify for probability vectors $x,y\in\mathbb R^n$ (e.g., containing the spectrum of any two quantum states), that is, for vectors $x,y\geq 0$ such that ${\bf e}^\top x=
{\bf e}^\top y=1$: in this case the ``thermomajorization curve''-criterion reduces to 
$\mathrm{th}_{d,x}(c)\leq\mathrm{th}_{d,y}(c)$ for all $c\in[0,{\bf e}^\top d]$ (resp.~all
$c\in\{\sum_{i=1}^j d_{\tau(i)}:j=1,\ldots,n-1\}$ where $\tau$ sorts
$\frac{x}{d}$ non-increasingly). Starting from thermomajorization curves 
there even exists an algorithm to find a Gibbs-stochastic matrix 
which implements the state transition in question \cite{Shiraishi20}.
Either way, the reason thermomajorization curves are equivalent to conditions using the $1$-norm 
is their fundamental link by means of the Legendre transformation\footnote{
Recall that given a convex function $f:D\to\mathbb R$ on a connected domain 
$D\subseteq\mathbb R$, its Legendre-Fenchel-transformation
$f^*:D^*\to\mathbb R$ is defined to be
$f^*(p):=\sup_{x\in D}\big(px-f(x)\big)$
for all $p\in D^*:=\{p\in\mathbb R\,:\,\sup_{x\in D}(px-f(x))<\infty\}$
\cite[Ch.~3.3]{Borwein06}.
}, denoted by $(\cdot)^*$: for all non-zero $d\geq 0$, all $y\in\mathbb R^n$, and all $t\in\mathbb R$ one has
\begin{equation}\label{eq:legendre}
2(-\mathrm{th}_{d,y})^*(t)={\bf e}^\top y+t({\bf e}^\top d)+\|y+td\|_1
\end{equation}
This readily follows from the definition of the Legendre transformation together with the fact that
$\mathrm{th}_{d,y+td}(c)=\mathrm{th}_{d,y}(c)+ct$ for all 
$c\in[0,{\bf e}^\top d]$
as well as the fact that the maximum of any thermomajorization curve 
equals the sum over all non-negative entries of the initial state 
\cite[Lemma 13 (iii)]{vomEnde22}.
With this the equivalence of the corresponding characterizations of $d$-majorization is due to the fact that the Legendre 
transformation in an involution which respects (more precisely: reverses) order
\cite[Thm.~4.2.1]{Borwein06}.
While these conditions are mathematically equivalent, in 
practice it is often easier to prove things using the curves $\mathrm{th}_{d,y}$.
This empirical observation will also be supported by the main part of this paper.

\begin{Remark}[Thermomajorization for Zero Temperature]
\label{rem_temp_zero}
While taking the limit $T\to 0^+$ of $\mathrm{th}_{d,y}$ is mathematically possible---after restricting the minimum in Eq.~\eqref{eq:thermo_curve_char} to those $i$ for which $d_i>0$---the resulting function does not characterize thermomajorization anymore.
The reason for this is that simply taking the limit $T\to 0^+$ of the Gibbs vector disregards important physical aspects; indeed, taking the temperature-zero limit of the $d$-majorization \textnormal{conditions} instead results in the correct and more restrictive notion of \textnormal{upper-triangular majorization} \cite{Narasimhachar15} (cf.~also p.9 in their supplemental material).
\end{Remark}

The importance of the $1$-norm conditions is that they lead
to a beautiful 
characterization of $M_d(y)$ from~\eqref{eq:def_M_d_y},
cf.~\cite[Thm.~10]{vomEnde22}:
\begin{equation}\label{eq:M_d_poly}
M_d(y)=\big\{x\in\mathbb R^n\,:\,{\bf e}^\top x=
{\bf e}^\top y\text{ and }\forall_{m\in\{0,1\}^n}\ m^\top x
\leq  \mathrm{th}_{d,y}(m^\top d) \big\}
\end{equation}
The geometric interpretation of each of these inequalities is that every binary vector
\mbox{$m\in\{0,1\}^n\setminus\{0,{\bf e}\}$}
is the normal vector to a halfspace which limits $M_d(y)$. The location of said 
halfspace is determined by the value $ \mathrm{th}_{d,y}(m^\top d)$ and thus by 
the thermomajorization curve. Note that the orientation of these halfspaces is 
universal, i.e.~they are independent of any of the system parameters;
subsequently, $y,d$ only influence the
\textit{location} of the faces. 
Another way of expressing this is to say that thermomajorization polytopes are 
obtained by shifting the faces of a classical majorization polytope.
Note that this can lead to the situation where some of these halfspace conditions become redundant.
This description of $M_d(y)$ has been used to prove continuity of 
the map $(d,P)\mapsto M_d(P)$ where $d>0$ and $P$ is from the collection of
non-empty compact sets in $\mathbb R^n$ equipped with the Hausdorff metric
\cite[Thm.~12~(ii)]{vomEnde22}. Alternatively, this result can be obtained
from continuity of the set of Gibbs-stochastic matrices in $H_0$ and $T>0$ \cite[Thm.~5.1]{Hartfiel74}.

Now, writing the bounded set $M_d(y)$ as the solution to finitely many inequalities shows that it 
is a convex polytope,
i.e.~it 
ultimately can be written as the convex hull of finitely many points
\cite{Gruenbaum03}.
Rather than just being an abstract
result, these extreme points have an analytic form.
Moreover, halfspaces becoming redundant leads to the coalescence of extreme points, more on this in Sec.~\ref{sec_degen}.
\subsection{Transportation Polytopes}\label{sec_transp_pol}
It turns out that Gibbs-stochastic matrices are closely related to transportation matrices which sit at the heart of the well-studied field of transportation theory \cite{Klee68,Bolker72,Dubois73,Brualdi06}.
It appears that, so far, this has been very much overlooked: this notion does not even appear in the standard work on majorization by Marshall et al.~\cite{MarshallOlkin}, and the only papers from the quantum thermodynamics literature (that we are aware of) which used results from transportation theory are ones by Mazurek et al.~\cite{Mazurek18,Mazurek19}.

Assuming finite domains, \textit{transportation matrices} are non-negative matrices with fixed column- and row-sums. More precisely, these are matrices
$A\in\mathbb R_+^{m\times n}$ such that $A{\bf e}=r$ and $c^\top={\bf e}^\top A$ for some $r\in\mathbb R^m$, $c\in\mathbb R^n$ with ${\bf e}^\top r={\bf e}^\top c$.
The collection of all such matrices is called \textit{transportation polytope} and is often denoted by $\mathrm T(r,c)$.
As already observed by Hartfiel \cite{Hartfiel74}, the connection to our setting then is the following:
For non-zero temperatures (i.e.~$d>0$) there is a one-to-one correspondence between (the extreme points of) the Gibbs-stochastic matrices and (the extreme points of) the symmetric transportation polytope $\mathrm T(d,d)$ by means of the
isomorphism $X\mapsto X\operatorname{diag}(d)$.
From our point of view, pursuing this approach may seem counter-intuitive because the geometry of the
Gibbs-stochastic matrices is known to be more 
complicated than 
the thermomajorization polytope.
Already in three dimensions the number of extreme points of the Gibbs-stochastic matrices depends on the temperature of the bath \cite[Fig.~1]{Mazurek18}),
that is, on certain relations between the entries of $d$ \cite[App.~A]{vomEnde22}.
However, drawing this connection grants access to powerful tools from combinatorics and graph theory. Roughly speaking there is a relation between the extreme points of $\mathrm T(d,d)$ and spanning trees of associated bipartite graphs.
We need not go into too much detail on the underlying techniques (instead, we refer to \cite{Mazurek19}); rather we 
will adopt 
useful notions from this field and adapt them to thermomajorization 
as well as the associated polytope.
For this our starting point is a paper by Loewy et al.~\cite{Loewy91}
where conditions on the vector $d$ that classify
certain features of (the polytope of) Gibbs-stochastic matrices were identified.
We present these conditions---which Loewy et al.~simply called ``property (a)'' and ``property (b)''\footnote{
To be precise, while Definition~\ref{def_well_ord_stable}~(i) is the same as their ``property (a)'', Definition~\ref{def_well_ord_stable}~(ii) strengthens ``property (b)''.
However (i) and (ii) together are equivalent to property (a) and property (b) combined.
}---in the following definition:
\begin{Definition}\label{def_well_ord_stable}
Given $d\in\mathbb R_{++}^n$ define a map $\mathcal{D}:\mathcal P(\{1,\ldots,n\})\to[0,\infty)$ on the power set 
of $\{1,\ldots,n\}$ via $\mathcal{D}(I):=\sum_{i\in I}d_i$.
\begin{itemize}
\item[(i)] We say $d$ is \emph{well structured} if
for all $I,J\subseteq\{1,\ldots,n\}$, $|I|<|J|$ one has $\mathcal{D}(I)<\mathcal{D}(J)$.
\item[(ii)] We call $d$ \emph{stable} if
$\mathcal{D}$ is injective.
\end{itemize}
\end{Definition}
\noindent Some remarks on these notions are in order:
By definition $d$ is stable if summing up two sets of entries of $d$ only yields the same result if the entries coincided in the first place.
In particular, stability implies that $d$ is non-degenerate.
Note that for non-degenerate systems stability is a generic property as only finitely many temperatures give rise to unstable Gibbs states. 
On the other hand, $d$ is well structured if
summing up $k-1$ arbitrary entries of $d$ always yields less than summing up any $k$ entries of $d$.
Interestingly, this notion is fully captured by the inequality
\begin{equation}\label{eq:ineq_wellstr}
\sum_{i=1}^{\lceil\frac{n}{2}\rceil-1}d_i^\downarrow<\sum_{i=n-\lceil\frac{n}{2}\rceil+1}^{n}d_i^\downarrow
\end{equation}
where $d_i^\downarrow:=(d^\downarrow)_i$ is the $i$-th largest component of $d$, in the sense that $d$ is well structured if and only if~\eqref{eq:ineq_wellstr} holds.
One way to see this is to first reduce well-structuredness of $d$ to a family of inequalities $\sum_{i=1}^{\alpha}d_i^\downarrow<\sum_{i=n-\alpha}^{n}d_i^\downarrow$, $\alpha=1,\ldots,n-1$ (where $\alpha$ plays the role of $|I|$ from Def.~\ref{def_well_ord_stable}), and in a second step realize that the inequality corresponding to $\alpha=\lceil\frac{n}{2}\rceil-1$ implies all other ones.
Either way, from~\eqref{eq:ineq_wellstr} one sees that well-structuredness is a high-temperature phenomenon:
Given energies of the system $E_1\leq\ldots\leq E_n$ (w.l.o.g.~$E_1<E_n$ to avoid the trivial case of full degeneracy) there exists a unique critical temperature $T_c\geq 0$ such that
\mbox{$
\sum_{i=1}^{\lceil\frac{n}{2}\rceil-1}e^{-E_i/T_c}=\sum_{i=n-\lceil\frac{n}{2}\rceil+1}^{n}e^{-E_i/T_c}
$},
and the corresponding Gibbs-vector is well structured
if and only if $T> T_c$.
One way to prove this is to examine the auxiliary function $\phi:\mathbb R_{+}\to \mathbb R_{+}$ defined via
$$
\phi(T):=\frac{\sum_{i=n-\lceil\frac{n}{2}\rceil+1}^{n}e^{-E_i/T}}{\sum_{i=1}^{\lceil\frac{n}{2}\rceil-1}e^{-E_i/T}}
$$
and to see that $\lim_{T\to 0^+}\phi(T)\in[0,1]$, $\lim_{T\to\infty}\phi(T)>1$, and $\phi'(T)>0$ for all\footnote{
This follows at once from the readily verified expression
$$
\phi'(T)=\Big(T\sum_{i=1}^{\lceil\frac{n}{2}\rceil-1}e^{-E_i/T}\Big)^{-2}\sum_{i=n-\lceil\frac{n}{2}\rceil+1}^{n}\sum_{j=1}^{\lceil\frac{n}{2}\rceil-1}(E_i-E_j)e^{-(E_i+E_j)/T}
$$
together with the observation that $E_i-E_j\geq 0$ because $i>j$, and even $E_n-E_1>0$ by assumption.
}
$T>0$. 
Thus by the intermediate value theorem there exists unique $T_c\geq0$ such that $\phi(T_c)=1$, and $\phi(T)>1$ (which is equivalent to~\eqref{eq:ineq_wellstr}) holds if and only if $T>T_c$. 

\begin{figure}[!ht]
\centering
\includegraphics[width=0.32\textwidth]{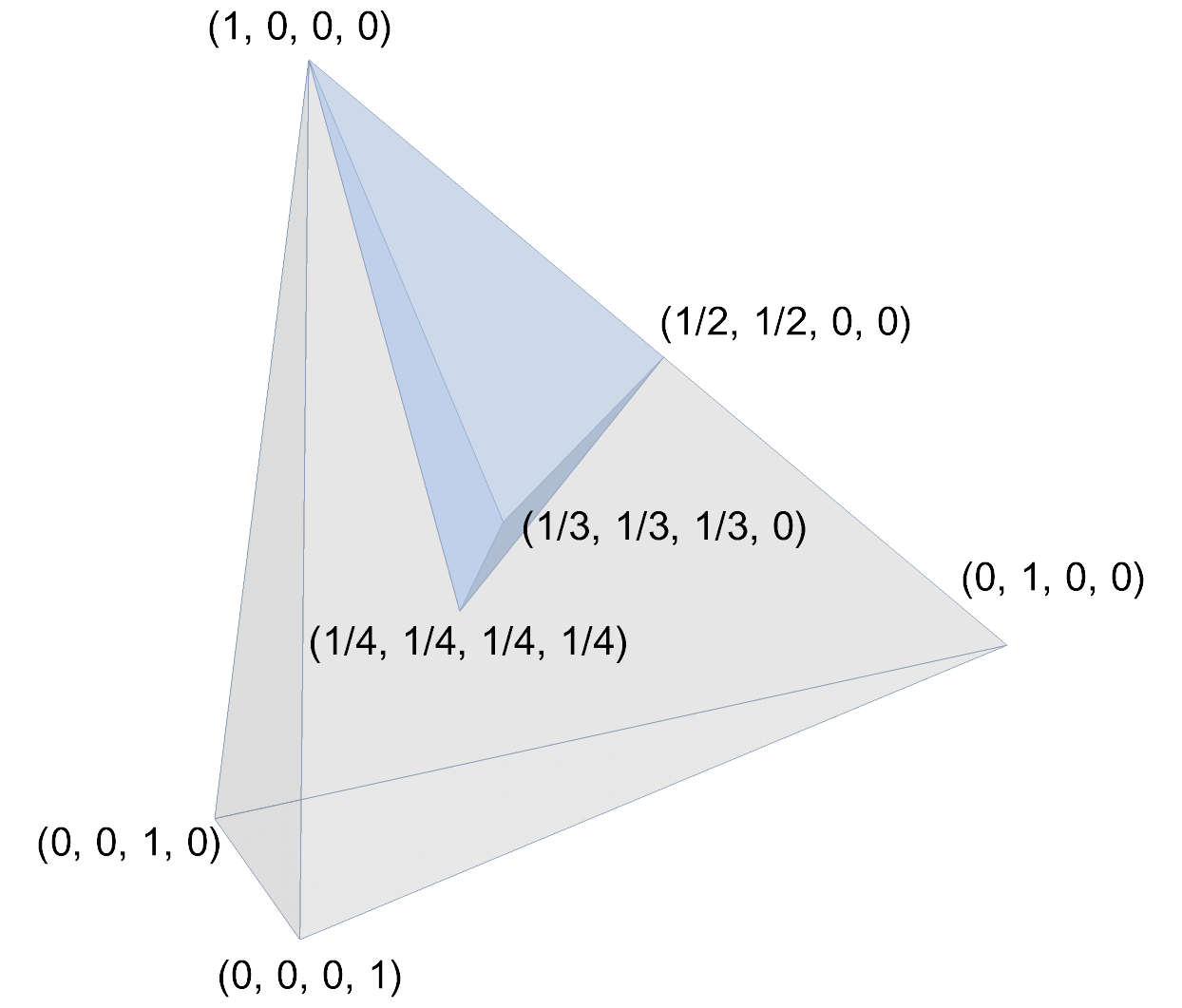}
\includegraphics[width=0.32\textwidth]{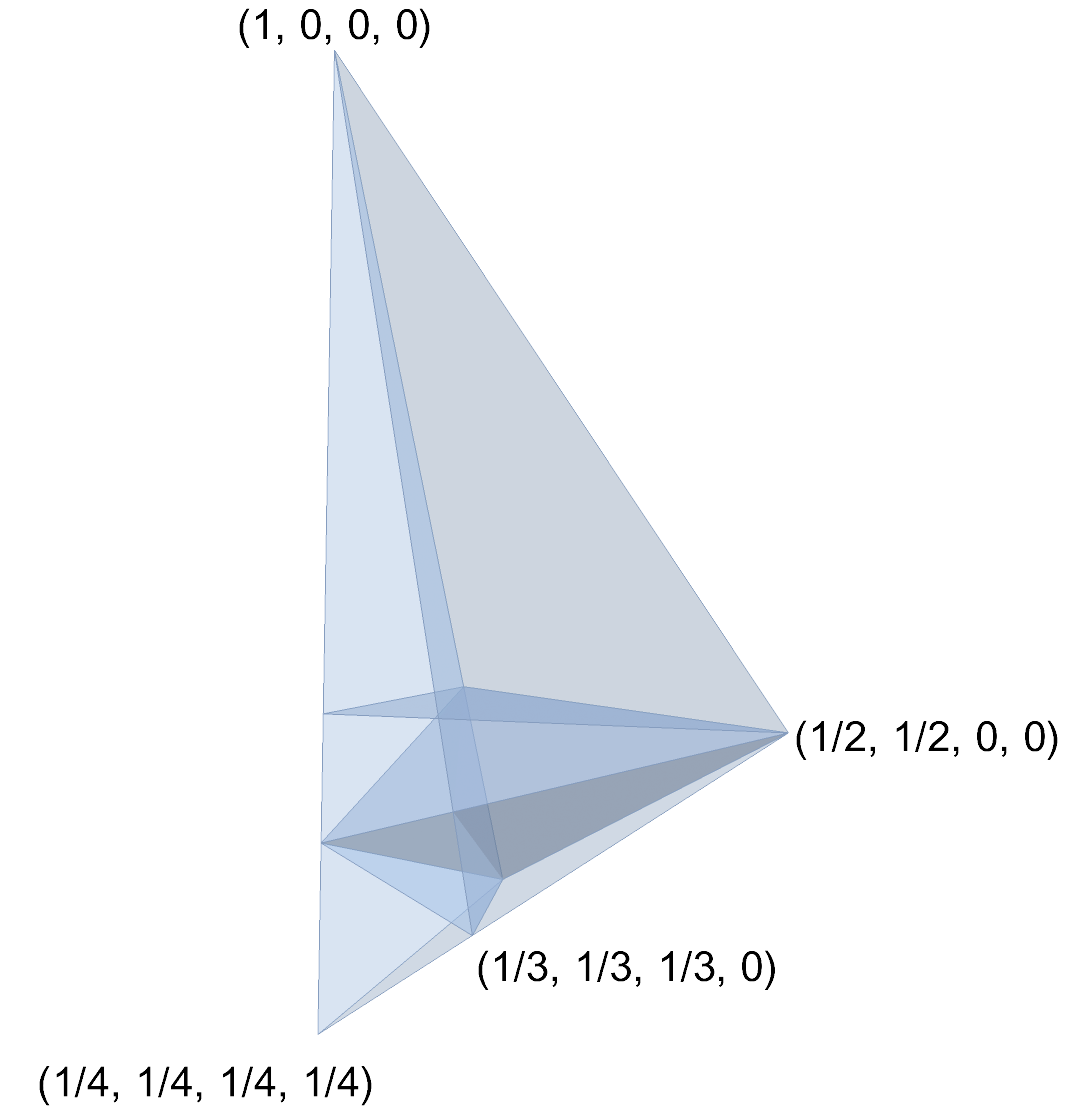}
\includegraphics[width=0.32\textwidth]{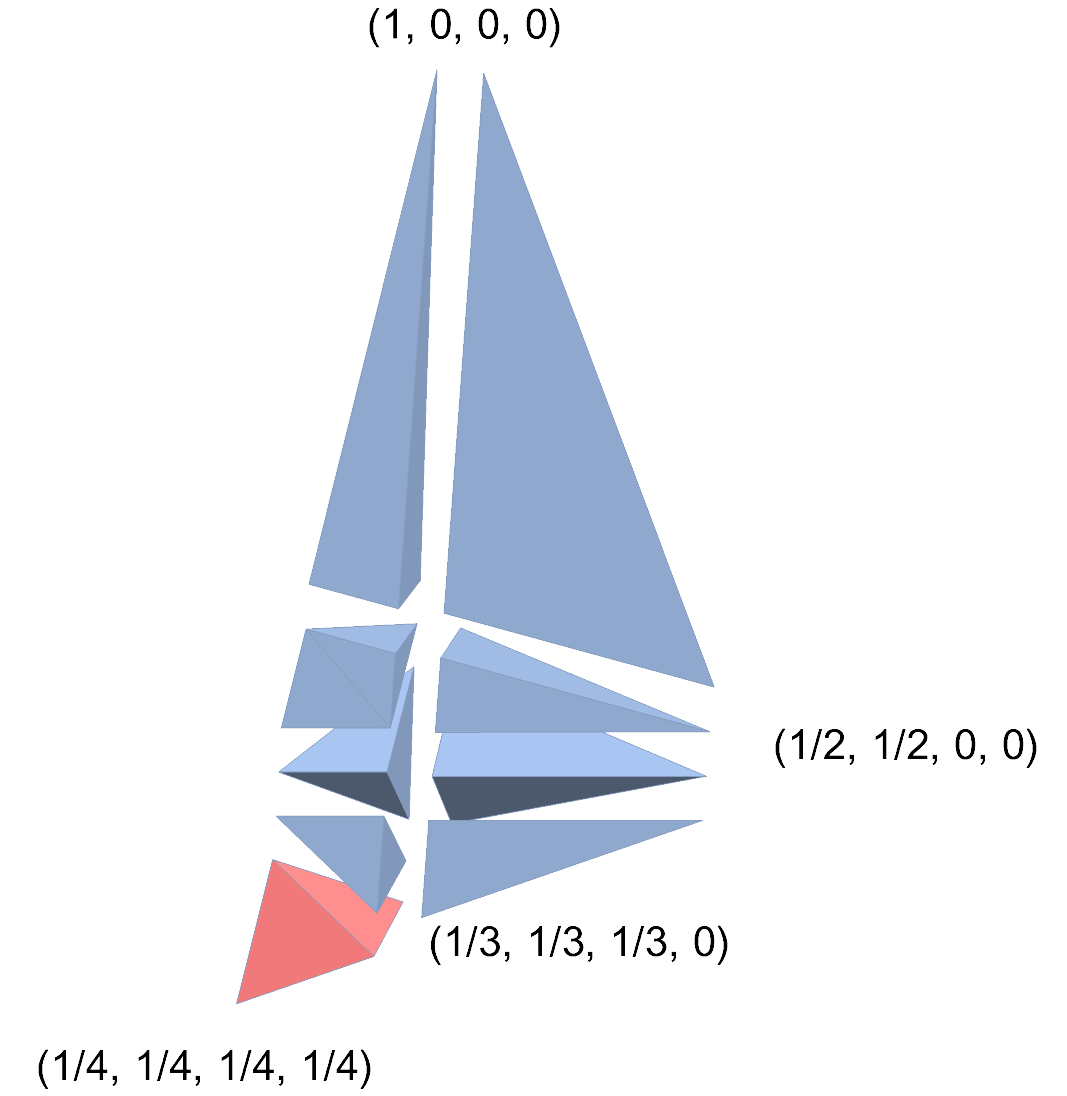}
\caption{Illustration of stability and well-structuredness. 
Consider $d\in\mathbb R^4_{++}$ where, w.l.o.g., ${\bf e}^\top d=1$.
Then $d$ lies in the relative interior of the standard simplex shown on the left.
By reordering its entries in a non-increasing fashion, cf.~Remark~\ref{rem_perm_matrix}, we can assume that $d$ lies in the ordered Weyl chamber shown in the middle.
The unstable points are composed of the walls of the Weyl chamber as well as five planes intersecting the Weyl chamber.
These planes cut the Weyl chamber into nine subchambers, and the one which includes the maximally mixed state ${\bf e}/4$
(highlighted in red on the right)
contains exactly the well-structured Gibbs vectors.}
\label{fig:subchambers}
\end{figure}

Either way the notion of stable Gibbs states as well as the fact that well-structuredness of the Gibbs state appears if (and only if) the temperature exceeds a critical value can be nicely visualized via the standard simplex and the ordered Weyl chamber, cf.~Figure \ref{fig:subchambers}.

Now based on these notions 
Loewy et al.~were able to prove the following:
Given any $d\in\mathbb R_{++}^n$ well-structuredness of $d$ is equivalent to every extreme point of $s_d(n)$ (i.e.~the Gibbs-stochastic matrices, recall footnote~\ref{footnote_gibbs_stoch_mat}) being invertible \cite[Thm.~3.1]{Loewy91}, the number of extreme points of $s_d(n)$ is maximal (when $d$ is taken as a parameter)
if and only if $d$ is well structured and stable \cite[Thm.~5.1 \& 6.1]{Loewy91},
and as a lower and upper bound to the number of extreme points of $s_d(n)$ they found $(n-1)!n^{n-2}$ and $n!n^{n-2}$, respectively\footnote{
Actually they explained how to calculate an even better lower bound which  grows asymptotically as $(n-1)!n^{n-2}\log n$
but cannot be written down as nicely as the bound $(n-1)!n^{n-2}$.
}.
The latter was a substantial improvement over the
lower bound $n!$ as first proven by Perfect and Mirsky~\cite{Perfect63}.

\section{Results}\label{sec_results}
As we will see the notions of well-structured and stable Gibbs states 
are key to answering fundamental questions in quantum thermodynamics.
Not only do this definition and the associated, already known results (e.g., on extreme points of the Gibbs-stochastic matrices) 
carry over
to our setting, this 
language is even suited to solve seemingly unrelated problems in quantum 
thermodynamics and, subsequently, lets us uncover new connections.
Consequently, this section will feature two types of results: first, ones which are at most superficially concerned with the geometry of the thermomajorization polytope (Sec.~\ref{sec_recap_2}), followed by some results on geometric quantities (e.g., extreme points) of said polytope, cf.~Sec.~\ref{sec_degen}.
The former can be seen as general principles underlying quantum thermodynamics
while the latter are more state-dependent and more explicit in nature.

\subsection{Cyclic State Transfers and Sub{space}s in Equilibrium}\label{sec_recap_2}

An overarching framework for this section is given by the notion of catalysis. There, in the most constrained form one calls a state transition $\rho\mapsto\omega$ \textit{strictly catalytic} if there exists an ancilla in state $\omega_C$ as well as an ``allowed'' operation $\Phi$ on the new overall system 
such that $\Phi(\rho\otimes\omega_C)=\omega\otimes\omega_C$ (although there are also ``weaker'' versions of catalytic transformations, cf.~the review article \cite{LBWN23}). 
For thermal operations in the quasi-classical realm strict catalysis boils down to state transfers $x\otimes z\prec_d y\otimes z$.

The idea behind such processes is of course that the catalyst $\omega_C$, resp.~$z$, undergoes a cyclic process in order to be returned (uncorrelated and) unchanged. 
This raises fundamental questions such as, e.g., what cyclic processes are even possible in our thermodynamic framework, which properties do such processes have, et cetera.
This is what our first result is about: in physical terms it states that \textit{global} cyclic thermodynamic processes are impossible for almost all temperatures (at least without access to external resources). 
While Thm.~\ref{thm_partial_order} below is concerned with two-step processes, as an immediate corollary one obtains an analogous result for time-continuous cyclic processes.
The precise statement here is that the impossibility of non-trivial cyclic processes is in one-to-one correspondence to
the notion of stable Gibbs vectors:
%
\begin{Theorem}\label{thm_partial_order}
Given $d\in\mathbb R_{++}^n$ the following statements are equivalent:
\begin{itemize}
\item[(i)] $d$ is stable.
\item[(ii)] Given any $x,y\in\mathbb R^n$, if $x\prec_dy\prec_dx$, then $x=y$.
\end{itemize}
\end{Theorem}
\begin{proof} 
``(ii) $\Rightarrow$ (i)'': We will prove this direction by contraposition, that is, given $d$ not stable we construct $x,y$ with $x\neq y$ such that $x\prec_d y\prec_dx$.
Indeed, assume to the contrary that there exist $I,J\subseteq\{1,\ldots,n\}$, $I\neq J$ such that $\sum_{i\in I}d_i=\sum_{j\in J}d_j$. Define
$x:=\sum_{i\in I}d_ie_i$, $y:=\sum_{j\in J}d_je_j$
and note that $I\neq J$ implies $x\neq y$.
We claim that $x\prec_dy\prec_dx$.
Recalling Sec.~\ref{sec_recap} we prove this, equivalently, by showing that $\|x-td\|_1=\|y-td\|_1$ for all $t\in\mathbb R$:
\begin{align*}
\|x-td\|_1&=\sum_{i\in I}|d_i-td_i|+\sum_{i\in \{1,\ldots,n\}\setminus I}|0-td_i|\\
&=\Big(\sum_{i\in I}d_i\Big)|1-t|+\Big({\bf e}^\top d-\sum_{i\in I}d_i\Big)|t|\\
&=\Big(\sum_{j\in J}d_j\Big)|1-t|+\Big({\bf e}^\top d-\sum_{j\in J}d_j\Big)|t|
=\ldots=\|y-td\|_1
\end{align*}

``(i) $\Rightarrow$ (ii)'':
The idea of this part of the proof is that any vectors 
$x\prec_dy\prec_dx$ induce the same thermomajorization curve, and---because $d$ is stable---applying this to the points where the curves change slope lets us conclude $x=y$.
More precisely, assume $d$ is stable and let $x,y\in\mathbb R^n$ be given such that 
$x\prec_d y\prec_d x$.
This implies ${\bf e}^\top x={\bf e}^\top y$ and, more importantly,
$\mathrm{th}_{d,x}(c)=\mathrm{th}_{d,y}(c)$ for all $c\in[0,{\bf e}^\top d]$. 
But these are piecewise linear functions characterized by its elbow points
so in particular the 
points where $\mathrm{th}_{d,x},\mathrm{th}_{d,y}$ have a (non-trivial) change in 
slope coincide (recall Sec.~\ref{sec_recap}).
To be more precise: there exists $k\in\{1,\ldots, n\}$ and sets $I^x_1,\ldots,I^x_{k-1},I^y_1,\ldots,I^y_{k-1}\in\mathcal P(\{1,\ldots,n\})$ such that
\begin{itemize}
\item $
\emptyset\neq I^x_1\subsetneq I^x_2\subsetneq \ldots\subsetneq I^x_{k-1}\subsetneq \{1,\ldots,n\}$ and $\emptyset\neq I^y_1\subsetneq I^y_2\subsetneq \ldots\subsetneq I^y_{k-1}\subsetneq \{1,\ldots,n\}
$.
\item $\mathrm{th}_{d,x}$ changes slope precisely at the 
inputs $\{\sum_{i\in I^x_l}d_i\,:\,l=1,\ldots,k-1\}$, and
$\mathrm{th}_{d,y}$ changes slope precisely at
$\{\sum_{i\in I^y_l}d_i\,:\,l=1,\ldots,k-1\}$.
\item $\mathrm{th}_{d,x}(\sum_{i\in I^x_l}d_i)=\sum_{i\in I^x_l}x_i$ and $\mathrm{th}_{d,y}(\sum_{i\in I^y_l}d_i)=\sum_{i\in I^y_l}y_i$ for all $l=1,\ldots,k-1$.
\end{itemize}
But because the changes in slope coincide, this (due to $d>0$) shows $\sum_{i\in I^x_l}d_i=\sum_{i\in I^y_l}d_i$.
By assumption $d$ is stable so we obtain $I^x_l=I^y_l$ for all $l=1,\ldots,k-1$. In particular
$$
\sum_{i\in I^x_l}x_i=\mathrm{th}_{d,x}\Big(\sum_{i\in I^x_l}d_i\Big)=
\mathrm{th}_{d,y}\Big(\sum_{i\in I^x_l}d_i\Big)=
\mathrm{th}_{d,y}\Big(\sum_{i\in I^y_l}d_i\Big)=
\sum_{i\in I^y_l}y_i=
\sum_{i\in I^x_l}y_i
$$
for all $l=0,\ldots,k$ when defining $I^x_0:=\emptyset=:I^y_0$ and $I^x_k:=\{1,\ldots,n\}=:I^y_k$.
Now that we took care of the points where the curves change slope all that remains are the points in between.
Consider any $l=1,\ldots,k$. Because $\mathrm{th}_{d,y}$ is affine linear on $[\sum_{i\in I^x_{l-1}}d_i,\sum_{i\in I^x_l}d_i  ]$ and the slope of $\mathrm{th}_{d,y}$ at ``the'' increment $d_i$ is given by $\frac{y_i}{d_i}$, there exists $c_{y,l}\in\mathbb R$ such that 
$y_i=c_{y,l}\cdot d_i$ for all $i\in I^x_l\setminus I^x_{l-1}$;
one argues analogously for $\mathrm{th}_{d,x}$ and obtains a constant $c_{x.l}$.
Be aware that we used stability of $d$ here: the length $\sum_{i\in I^x_l\setminus I^x_{l-1}}d_i$ of the interval $[\sum_{i\in I^x_{l-1}}d_i,\sum_{i\in I^x_l}d_i  ]$ can only come from adding up $\{d_i:i\in  I^x_l\setminus I^x_{l-1}\}$.
Now because $\mathrm{th}_{d,x}$ and $\mathrm{th}_{d,y}$ coincide, by assumption we find that $c_{x,l}=c_{y,l}(=:c_l)$ and thus
\begin{equation*}
x=\sum_{i=1}^n x_ie_i=\sum_{l=1}^k\sum_{i\in I^x_l\setminus I^x_{l-1}}x_ie_i=\sum_{l=1}^k\sum_{i\in I^x_l\setminus I^x_{l-1}}\underbrace{c_ld_i}_{=y_i}e_i=y.\tag*{\qedhere}
\end{equation*}
\end{proof}

Of course this result does not prohibit \textit{local} cyclic processes, that is, thermodynamic processes where only a subsystem returns to its original state at the end (so precisely, catalysis).
What Thm.~\ref{thm_partial_order} does assert, however, is that in general a quasi-classical cyclic process (modeled by thermal operations) which is \textit{not} local has to use up some external resource along the way.\medskip


{
For the remainder of this section our focus lies on (states on) sub{space}s which are ``in equilibrium''.
The motivation behind this notion is that if a state restricted to some subspace is a multiple of the Gibbs state, then all
thermal operations act trivially on it: Indeed given a state $x$ and a subspace $P$ such that $x|_P$ is a multiple of the Gibbs vector, then any Gibbs-stochastic matrix which acts non-trivially only on $P$---i.e.~it is of the form $A=A_P\oplus{\bf1}_{P^\perp}$---necessarily leaves $x$ invariant. 
}
This of course generalizes to subsystems of coupled systems by choosing $P$ 
appropriately.
The precise definition reads as follows:
\begin{Definition} \label{def_subsys_equi}
Let $d\in\mathbb R_{++}^n$, $y\in\mathbb R^n$. 
We say that a subset $P\subseteq\{1,\ldots,n\}$, $|P|>1$ of the system's energy levels is \emph{in equilibrium} if $\frac{y_i}{d_i}=\frac{y_j}{d_j}$ for all $i,j\in P$.
On the other hand if no such subset $P$ satisfies this condition we say that the system is in \emph{total non-equilibrium}.
\end{Definition}


In this language our next result states that regardless of whether {there is a subspace} in equilibrium (as long as the full system is not) every such sub{space} can be brought out of equilibrium by means of $d$-stochastic matrices.
This in particular applies to catalytic state transfers: if, for example, a system is in equilibrium, then any catalyst (which itself is not in the Gibbs state) allows for bringing arbitrary energy levels of the original system out of equilibrium. 
The precise statement is derived via the dimension of the thermomajorization polytope and reads as follows:

\begin{Theorem}\label{thm_polytope_dimension}
Let $d\in\mathbb R_{++}^n$, $y\in\mathbb R^n$. The following statements are equivalent.
\begin{itemize}
\item[(i)] $M_d(y)$ is not singular, that is, $M_d(y)$ consists of more than just $y$.
\item[(ii)] $y$ is not a multiple of $d$.
\item[(iii)] The dimension of the convex polytope $M_d(y)$ is maximal, i.e.~its dimension is $n-1$ which is equal to the dimension of the standard simplex $\Delta^{n-1}$ of all $n$-dimensional probability vectors.
\end{itemize}
In particular, if there exists a subset $P\subsetneq\{1,\ldots,n\}$, $|P|>1$ in equilibrium (i.e.~${y_i}/{d_i}={y_j}/{d_j}$ for all $i,j\in P$), then there exists $z\in M_d(y)$ such that $z$ is in total non-equilibrium.
\end{Theorem}
\begin{proof}
``(iii) $\Rightarrow$ (i)'': Trivial.
``(i) $\Rightarrow$ (ii)'': Obvious via contrapositive.
``(ii) $\Rightarrow$ (iii)'':
The dimension of $M_d(y)$ is trivially upper bounded by $n-1$
as it is a subset of the $n-1$-dimensional standard simplex \cite[Coro.~17]{vomEnde22}. For the converse we argue by contraposition: if the dimension is strictly less than $n-1$, then there
must exist a condition in \eqref{eq:M_d_poly} which
is an equality \cite[Ch.~8.2]{Schrijver86}.
More precisely, there must exist
$m\in\{0,1\}^n$, $0<{\bf e}^\top m<n$ and $c\in\mathbb R$ such that 
$m^\top x=c$
for all $x\in M_d(y)$.
First we determine $c$. Let $\sigma\in S_n$ be any permutation such that $\underline{\sigma}m$ is sorted non-increasingly, i.e.~$\underline{\sigma}m=(1,\ldots,1,0,\ldots,0)^\top$. We know $E_{d,y}(\sigma)\in M_d(y)$ so
\begin{align*}
c=m^\top E_{d,y}(\sigma)
=\sum_{j=1}^{{\bf e}^\top m}(E_{d,y}(\sigma))_{\sigma(j)}=\mathrm{th}_{d,y}\Big(\sum_{j=1}^{{\bf e}^\top m}d_{\sigma(j)}\Big)=\mathrm{th}_{d,y}(m^\top d)\,.
\end{align*}
The final step is to evaluate the condition $m^\top x=\mathrm{th}_{d,y}(m^\top d)$ at a multiple of $d$:
because $\frac{d{\bf e}^\top}{{\bf e}^\top d}\in s_d(n)$ one has $\frac{d{\bf e}^\top}{{\bf e}^\top d}y=\frac{{\bf e}^\top y}{{\bf e}^\top d}d\in M_d(y)$. Therefore
\begin{align}
\mathrm{th}_{d,y}(m^\top d)=c=m^\top \Big(\frac{{\bf e}^\top y}{{\bf e}^\top d}d\Big)&=\frac{{\bf e}^\top d-m^\top d}{{\bf e}^\top d}\cdot 0+\frac{m^\top d}{{\bf e}^\top d}{\bf e}^\top y\notag\\
&=\frac{{\bf e}^\top d-m^\top d}{{\bf e}^\top d-0}\mathrm{th}_{d,y}(0)+\frac{m^\top d-0}{{\bf e}^\top d-0}\mathrm{th}_{d,y}({\bf e}^\top d)\,.
\label{eq:th_d_y_affine_linear}
\end{align}
Because $\mathrm{th}_{d,y}$ is concave and because $m^\top d\in (0,{\bf e}^\top d)$ by the assumptions on $m$ and $d$,
Lemma~\ref{lemma_concave_connecting_line} (iv) (App.~\ref{app_concave_lemma}) shows that \eqref{eq:th_d_y_affine_linear}
can only hold if $\mathrm{th}_{d,y}$ is linear. But the latter is equivalent to $y$ being a multiple of $d$ as,
by definition, the slopes of $\mathrm{th}_{d,y}$ are given by ${y_j}/{d_j}$. 

Finally, the additional statement follows at once from the following two facts: (a) The collection of all vectors in total non-equilibrium is dense in $\Delta^{n-1}$, and (b) $M_d(y)$ contains an interior point w.r.t.~the hyperplane $W_{{\bf e}}:=\{z\in\mathbb R^n:{{\bf e}}^\top z=1\}$.
While (b) is due to (iii), for (a)
note that given any $P\subsetneq\{1,\ldots,n\}$, $|P|>1$ the set of vectors $y\in W_{{\bf e}}$ in equilibrium (w.r.t.~$P$) form a lower-dimensional subspace of $W_{{\bf e}}$. In particular this set is nowhere dense\footnote{
Recall that a subset of a topological space is called \textit{nowhere dense} if
its closure has empty interior \cite{Willard70}.
},
which continues to hold when taking the union over all (finitely many!) such $P$.
But this, in particular, implies that the complement of this set---that is, the collection of all vectors in total non-equilibrium---is dense in $W_{{\bf e}}$. This concludes the proof.
\end{proof}

\noindent Returning to the language of sub{space}s in equilibrium note that this result is an existence result and does not infer anything about the potential ``amplitude'' of such transfers.
Mathematically, Thm.~\ref{thm_polytope_dimension} complements the old result of Hartfiel that the dimension of $s_d(n)$ for all $d>0$ is $(n-1)^2$ \cite[Thm.~3.1]{Hartfiel74}.

\begin{Remark}
The assumption $d>0$ in Thm.~\ref{thm_polytope_dimension} is necessary as hinted at by the fact that $\operatorname{dim}(s_d(n))$ for general $d\in\mathbb R_+^n$ depends on the number of zeros in $d$
\cite[Thm.~3.2]{Hartfiel74}.
This is nicely illustrated by the simple example $d=(1,1,0)^\top$, $y=(1,0,0)^\top$ because then $M_d(y)=\{(c,1-c,0)^\top\,:\,c\in[0,1]\}$ which is not two-, but only one-dimensional.
\end{Remark}


\subsection{Extreme Points of the Thermomajorization Polytope} \label{sec_degen}
{Let us stress that---until now---the concept of subspaces in equilibrium (and hence Thm.~\ref{thm_polytope_dimension}) is logically independent from the notions of stability and well-structuredness.
This missing connection will be established below, where
well-structured states and sub{space}s in equilibrium will be linked via geometric properties of the thermomajorization polytope, in particular the number of its extreme points.
The key mathematical object for doing so}
is the extreme point map $E_{d,y}$ which is defined as follows:\footnote{An analytic form of the extreme points of $M_d(y)$ has appeared independently in the physics~\cite{Lostaglio18,Alhambra19} and the mathematics~\cite{vomEnde22} literature.}

\begin{Definition} \label{def_th_E}
Let $d\in\mathbb R_{++}^n$, $y\in\mathbb R^n$. Define the 
\emph{extreme point map} $E_{d,y}$ on the symmetric group $S_n$ via
\begin{equation*}
E_{d,y}:S_n\to\mathbb R^n,\quad
\sigma\mapsto\Big( \mathrm{th}_{d,y}\Big( \sum_{i=1}^{\sigma^{-1}(j)}
d_{\sigma(i)} \Big)-\mathrm{th}_{d,y}\Big( \sum_{i=1}^{\sigma^{-1}(j)-1}
d_{\sigma(i)} \Big) \Big)_{j=1}^n\,.
\end{equation*}
Equivalently,
$(\underline{\sigma}E_{d,y}(\sigma))_j=(E_{d,y}(\sigma))_{\sigma(j)}=
\mathrm{th}_{d,y}\big( \sum_{i=1}^{j}d_{\sigma(i)} \big)-\mathrm{th}_{d,y}
\big( \sum_{i=1}^{j-1}d_{\sigma(i)} \big)$,
where $\underline{\sigma}$ is the permutation matrix corresponding to
$\sigma$ as defined above (Remark~\ref{rem_perm_matrix}).
\end{Definition}

As the name suggests, for all $d\in\mathbb R_{++}^n$ and $y\in\mathbb R^n$ the image of $E_{d,y}$ equals the set $\mathrm{ext}(M_d(y))$ of extreme points of $M_d(y)$,
and thus $M_d(y)=\operatorname{conv}\{E_{d,y}(\sigma):\sigma\in S_n\}$
\cite[Thm.~16]{vomEnde22}.
It should be noted that the extreme point property also manifests in the thermomajorization curves, relating to the concept of \textit{tight thermomajorization}:
Given any $y,z\in\mathbb R_+^n$, the point $z$ is an extreme point of $M_d(y)$ if and only if all elbow points of $\mathrm{th}_{d,z}$ lie on the curve $\mathrm{th}_{d,y}$ \cite[Thm.~2]{Mazurek19}.
Also there is a nice connection between the extreme points of $M_d(y)$ and a special class of extreme points of the Gibbs-stochastic matrices which we will elaborate on at the end of this section.
Moreover, Sec.~\ref{sec_Mdy} below presents a step-by-step calculation of the extreme point map and explains how it, equivalently, can be computed graphically using the
thermomajorization curve, cf.~Figure~\ref{fig_thermocurve_ex_1}.

Our focus for now, however, lies on properties of the map $E_{d,y}$ and geometric aspects of $M_d(y)$. From Def.~\ref{def_th_E} it is clear that the maximal number of extreme points of $M_d(y)$ is $n!=|S_n|$; if there are strictly fewer than $n!$ extreme points we say that the polytope is \textit{degenerate}.
The goal of this section is to prove the following result which 
states that degeneracy of the polytope is a ``witness'' for sub{space}s in equilibrium, assuming $d$ is well structured (equivalently: assuming large enough temperatures, cf.~Sec.~\ref{sec_transp_pol}).

\begin{Theorem} 
\label{thm_necessary_degen}
Let $d\in\mathbb R_{++}^n$, $y\in\mathbb R^n$. 
If $E_{d,y}$ is not injective, i.e.~$|\mathrm{ext}(M_d(y))|<n!$, then at least one of the following holds:
\begin{itemize}
\item[(i)] $y$ has a sub{space} which is in equilibrium with respect to $d$.
\item[(ii)] $d$ is not well structured, i.e.~the temperature is below the critical value $T\leq T_c$, cf.~Sec.~\ref{sec_transp_pol}.
\end{itemize}
\end{Theorem}

Note that in the example of Sec.~\ref{sec_Mdy} degenerate extreme points occur, and the degeneracy stems from the fact that there exists a sub{space} in equilibrium.
This 
can be shown generally 
as specifying the preimage of $y$ under the extreme point map $E_{d,y}$ turns out to be straightforward:


\begin{Lemma}\label{lem_y_deg}
Let $d\in\mathbb R_{++}^n$, $y\in\mathbb R^n$, and $\sigma\in S_n$ be given. One has $E_{d,y}(\sigma)=y$ if and only if $\frac{y_{\sigma(1)}}{d_{\sigma(1)}}\geq\ldots\geq\frac{y_{\sigma(n)}}{d_{\sigma(n)}}$.
\end{Lemma}

\begin{proof}
Assume $\sigma\in S_n$ satisfies $E_{d,y}(\sigma)=y$ so $(E_{d,y}(\sigma))_{\sigma(j)}=y_{\sigma(j)}$ for all $j=1,\ldots,n$. By definition of $E_{d,y}$ this means $\mathrm{th}_{d,y}(\sum_{i=1}^j d_{\sigma(i)})=\mathrm{th}_{d,y}(\sum_{i=1}^{j-1} d_{\sigma(i)})+y_{\sigma(j)}$ which by induction is equivalent to $\mathrm{th}_{d,y}(\sum_{i=1}^jd_{\sigma(i)})=\sum_{i=1}^j y_{\sigma(i)}$ for all $j=1,\ldots,n-1$. But by Lemma~\ref{lemma_th_ordered} (App.~\ref{app_concave_lemma}) this holds if and only if ${y_{\sigma(1)}}/{d_{\sigma(1)}}\geq\ldots\geq{y_{\sigma(n)}}/{d_{\sigma(n)}}$.
\end{proof}

\noindent Clearly then the extreme point $y$ is degenerate (in the sense that it has multiple preimages under $E_{d,y}$) if and only if there exists a sub{space} which is in equilibrium, cf.~Definition~\ref{def_subsys_equi}. 
The converse, however, is not true. 
Indeed, the example in Sec.~\ref{sec_degen_ex} shows that degenerate extreme points can occur even if the system is in total non-equilibrium.
Note however that in this example the vector $d$ is not well structured, indicating a low temperature, as required by Thm.~\ref{thm_necessary_degen}.

Note that Lemma~\ref{lem_y_deg} relates to the concept of virtual temperatures \cite{Janzing00,Singh21,Skrzypczyk15,Brunner12}
as multiple permutations are mapped to $y$ under $E_{d,y}$ if and only if there exist $i,j\in\{1,\ldots,n\}$ such that 
the background temperature equals
$
T=	\frac{E_j-E_i}{\ln(y_i)-\ln(y_j)}=:T_{ij}
$ (which is equivalent to $\frac{y_i}{d_i}=\frac{y_j}{d_j}$). 
In other words virtual temperatures characterize when
another corner of the polytope ``crosses'' the initial state.
This also relates to the notion of passivity: 
The degeneracy of $M_d(y)$ at temperature $T_{ij}$ corresponding to the transition between $E_i$ and $E_j$
is physical (i.e.~$T_{ij}>0$)
only if the initial state $\operatorname{diag}(y)$ is passive, meaning no work can be extracted via unitary transformations \cite[Sec.~III]{Skrzypczyk15}.

\begin{Example}
It is worth addressing the case of classical majorization, i.e.~$d={\bf e}$ (up to a factor which is of no consequence).
We find that
$\mathrm{th}_{d,y}(\sum_{i=1}^j d_{\sigma(i)})=\sum_{i=1}^j y_i^\downarrow$ 
for any $j=1,\ldots,n$, $\sigma\in S_n$;
this recovers
$E_{d,y}(\sigma)=\underline{\sigma}^{-1} y^\downarrow$ \cite[Ch.~4, Prop.~C.1]{MarshallOlkin}.
Therefore
degeneracies of 
$M_{{\bf e}}(y)$
correspond to some entries of the initial state coinciding\footnote{
Given any $y\in\mathbb R^n$ the number of extreme points of $M_{{\bf e}}(y)$ equals 
${n!}/({p_1!\cdot\ldots\cdot p_m!})$
where 
$(p_i)_{i=1}^m$ denote the multiplicities of 
$y$, that is, $y^\downarrow=\sum_{i=1}^m y_i^\downarrow\sum_{j=1}^{p_i}e_{p_1+\ldots+p_{i-1}+j}$ where $y_1^\downarrow>\ldots>y_m^\downarrow$.
}.
This, in fact, implies ``uniformity'' of the classical majorization polytope's degeneracy in the sense that the preimage of \textnormal{each} extreme point under $E_{d,y}$
has the same size, which is certainly false for general $d\in\mathbb R_{++}^n$, cf.~Sec.~\ref{sec_Mdy} and in particular Table~\ref{table1}.
This uniformity has to do with the fact that $s_{{\bf e}}(n)$ contains all permutation matrices which yields a group action of $S_n$ on $s_{{\bf e}}(n)$.
This group action is transitive on the vertices, and hence all vertices are equivalent.
This does not hold for general $d\in\mathbb R_{++}^n$: the inverse of some invertible element of $s_d(n)$ is again in $s_d(n)$ if and only if it is a permutation matrix \cite[Remark~4.1]{Joe90}.
\end{Example}

Going through the construction of extreme points in Sec.~\ref{sec_Mdy}, and especially the graphical approach, one notices quickly how degeneracies can occur, cf.~in particular Remark~\ref{rem_thermo_curves_interpretation}.
Indeed, given $d\in\mathbb R_{++}^n$, $y\in\mathbb R^n$ there exists $m\in\{0,\ldots,n-1\}$ as well as real numbers $\Delta_0,\ldots,\Delta_{m+1}$ such that $0=\Delta_0<\Delta_1<\ldots<\Delta_m<\Delta_{m+1}={\bf e}^\top d$ and $\mathrm{th}_{d,y}$ changes slope at an input $x\in(0,{\bf e}^\top d)$ if and only if $x=\Delta_k$ for some $k=1,\ldots,m$. 
The example in Sec.~\ref{sec_Mdy} shows that degeneracies occur when two or more intervals of length $d_i$, $i=1,\ldots,m$ (ordered according to $\sigma$) are contained in the same interval $(\Delta_{k-1},\Delta_k)$.

This motivates the following definition. For $\sigma\in S_n$, $k\in\{1,\ldots,m+1\}$ define
\begin{equation} \label{eq:intersection-index}
I_k^\sigma:=\Big\{j\in\{1,\ldots,n\}\,:\,\Big[ \sum_{i=1}^{\sigma^{-1}(j)-1}d_{\sigma(i)},\sum_{i=1}^{\sigma^{-1}(j)}d_{\sigma(i)} \Big]\cap(\Delta_{k-1},\Delta_k)\neq\emptyset\Big\}\,.
\end{equation}
The elements of $I_k^\sigma$ are those $j\in\{1,\ldots,n\}$ for which the interval of length $d_j$ corresponding to the partition induced by $\sigma$ intersects $(\Delta_{k-1},\Delta_k)$.
In particular 
$\bigcup_{k=1}^{m+1}I_k^\sigma=\{1,\ldots,n\}$ for all $\sigma\in S_n$.
An illustrative example is given in Sec.~\ref{sec_intersections}.

The importance of this definition
comes from its ability to characterize the image of the extreme point map
as the following result shows.

\begin{Proposition}\label{prop_I_sigma_coincide}
Let $d\in\mathbb R_{++}^n$, $y\in\mathbb R^n$ and let $m\in\{1,\ldots,n-1\}$ be the number of changes in slope of $\mathrm{th}_{d,y}$.
Given any $\sigma,\tau\in S_n$ one has $E_{d,y}(\sigma)=E_{d,y}(\tau)$ if and only if $I_k^\sigma=I_k^\tau$ for all $k=1,\ldots,m+1$.
\end{Proposition}
\begin{proof}
``$\Rightarrow$'':
Assume $E_{d,y}(\sigma)=E_{d,y}(\tau)$. Then
\begin{equation}\label{eq:sigma_tau_interval}
\mathrm{th}_{d,y}\Big( \sum_{i=1}^{\sigma^{-1}(j)}d_{\sigma(i)} \Big)-
\mathrm{th}_{d,y}
\Big( \sum_{i=1}^{\sigma^{-1}(j)-1}d_{\sigma(i)} \Big)=
\mathrm{th}_{d,y}\Big( \sum_{i=1}^{\tau^{-1}(j)}d_{\tau(i)} \Big)-
\mathrm{th}_{d,y}
\Big( \sum_{i=1}^{\tau^{-1}(j)-1}d_{\tau(i)} \Big)
\end{equation}
for all $j=1,\ldots,n$ Definition~\ref{def_th_E}.
Now given $j\in\{1,\ldots,n\}$ there are two (non-exclusive) possibilities: either $[\sum_{i=1}^{\sigma^{-1}(j)-1}d_{\sigma(i)},\allowbreak \sum_{i=1}^{\sigma^{-1}(j)}d_{\sigma(i)}] = [\sum_{i=1}^{\tau^{-1}(j)-1}d_{\tau(i)},\allowbreak \sum_{i=1}^{\tau^{-1}(j)}d_{\tau(i)}]$
which implies $j\in I_k^\sigma$ if and only if $j\in I_k^\tau$, or
the two intervals do not coincide. The latter implies that $\mathrm{th}_{d,y}$ is affine linear on
$
\operatorname{conv}([ \sum_{i=1}^{\sigma^{-1}(j)-1}d_{\sigma(i)},\sum_{i=1}^{\sigma^{-1}(j)}d_{\sigma(i)}]\cup[ \sum_{i=1}^{\tau^{-1}(j)-1}d_{\tau(i)},\sum_{i=1}^{\tau^{-1}(j)}d_{\tau(i)}] )
$
by
our argument from above.
But this means that both these intervals have to be contained in the same interval $[\Delta_{k-1},\Delta_k]$,
hence $j\in I_k^\sigma$ and $j\in I_k^\tau$. 

``$\Leftarrow$'':
Assume by contraposition that $E_{d,y}(\sigma)\neq E_{d,y}(\tau)$ so there exists $j=1,\ldots,n$ such that \eqref{eq:sigma_tau_interval} does not hold. 
Thus $J_\sigma:=[ \sum_{i=1}^{\sigma^{-1}(j)-1}d_{\sigma(i)},\sum_{i=1}^{\sigma^{-1}(j)}d_{\sigma(i)} ]$ has to differ from $J_\tau:=[ \sum_{i=1}^{\tau^{-1}(j)-1}d_{\tau(i)},\sum_{i=1}^{\tau^{-1}(j)}d_{\tau(i)} ]$ 
(which is only possible if $\sum_{i=1}^{\sigma^{-1}(j)-1}d_{\sigma(i)}\neq \sum_{i=1}^{\tau^{-1}(j)-1}d_{\tau(i)}$ because the intervals have the same length).
Moreover, $\mathrm{th}_{d,y}$ cannot be affine linear on $\operatorname{conv}(J_\sigma\cup J_\tau)$ so there exists $k\in\{1,\ldots,m\}$ such that the change of slope
$\Delta_k$ lies in the interior of $\operatorname{conv}(J_\sigma\cup J_\tau)$.

\begin{figure}[!ht]
\includegraphics[width=0.95\textwidth]{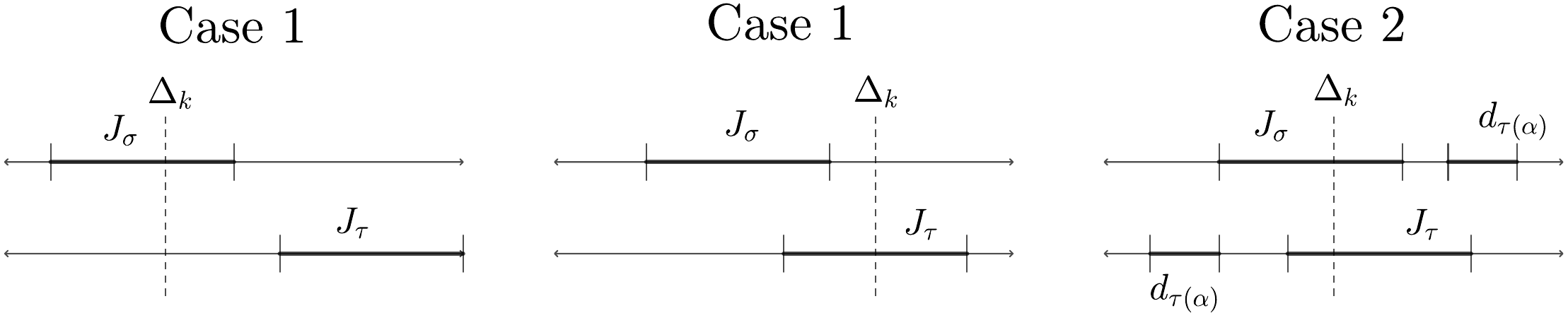}
\caption{Possible combinations of whether $J_\sigma$ and $J_\tau$ intersect, and where $\Delta_k$ lies relative to $J_\sigma,J_\tau$.}\label{figure_proof_prop3}
\end{figure}

Now there are two possible cases (cf.~Figure~\ref{figure_proof_prop3}): if $\Delta_k\not\in J_\sigma\cap J_\tau$, then $\Delta_k\not\in J_\sigma$ (or $\Delta_k\not\in J_\tau$) so $\operatorname{int}(J_\sigma)$ ($\operatorname{int}(J_\tau)$) is to the left or to the right of $\Delta_k$. Either way there exists $\tilde k\in\{1,\ldots,m+1\}$ such that $j\in I_{\tilde k}^\tau$ but $j\not\in I_{\tilde k}^\sigma$ so we are done.
For the second case assume that $\Delta_k\in J_\sigma\cap J_\tau$ and w.l.o.g.~that
$\min J_\sigma<\min J_\tau$, i.e.~$\sum_{i=1}^{\sigma^{-1}(j)-1}d_{\sigma(i)}< \sum_{i=1}^{\tau^{-1}(j)-1}d_{\tau(i)}$. Therefore one can find $\alpha\in\{1,\ldots,\tau^{-1}(j)-1\}$ such that $\tau(\alpha)\not\in\{\sigma(1),\ldots,\sigma(\sigma^{-1}(j)-1),j\}$. This in turn implies
$\sigma^{-1}(\tau(\alpha))>\sigma^{-1}(j)$, that is, the interval corresponding to
$d_{\tau(\alpha)}$ is to the left of $J_\tau\supseteq J_\tau\cap J_\sigma$ but to the right of $J_\sigma\supseteq J_\sigma\cap J_\tau$. Because $\Delta_k\in J_\sigma\cap J_\tau$ this yields $\tilde k\in\{1,\ldots,k\}$ such that $\tau(\alpha)\in I_{\tilde k}^\tau$ but $\tau(\alpha)=\sigma(\sigma^{-1}(\tau(\alpha)))\not\in I_{\tilde k}^\sigma$. This concludes the proof.
%
%
\end{proof}

\noindent This means that $\sigma$ and $\tau$ yield the same extreme point if and only if they differ exactly by a permutation of the intervals of length $d_i$ such that each interval remains in the same region where $\mathrm{th}_{d,y}$ is affine linear. 
Note that the case of classical majorization follows from this since in that case, such permutations differ exactly by elements of the stabilizer of $y$. Similarly,
the result about preimages of $y$ (Lemma~\ref{lem_y_deg}) follows from this.

Either way Prop.~\ref{prop_I_sigma_coincide} gives a simple criterion to check whether the image of different permutations under the extreme point map $E_{d,y}$ coincides or not. 
In particular, a bound on the degeneracy of any extreme point can be given by how many of the $d_i$ intervals fit into the same $[\Delta_{k-1},\Delta_k]$ interval. This is related to the bin packing problem in computer science which (is strongly NP-complete but) admits some reasonable approximations.
Another way to look at the above results is that $|\mathrm{ext}(M_d(y))| \leq n!+1-|(S_n)_{{y}/d}|$ by Lemma~\ref{lem_y_deg} where $(S_n)_{{y}/d}$ is the stabilizer of the vector ${y}/d=({y_i}/{d_i})_{i=1}^n$ in $S_n$, and the examples in Sec.~\ref{sec_Mdy} \&~\ref{sec_degen_ex} show that this bound is not tight. 
These examples suggest that improving this bound via an analytic expression is a non-trivial task. 
Finally, Prop.~\ref{prop_I_sigma_coincide} lets us prove Thm.~\ref{thm_necessary_degen}.
\begin{proof}[Proof of Thm.~\ref{thm_necessary_degen}]
Assume there exist distinct $\sigma,\tau\in S_n$ such that $E_{d,y}(\sigma)=E_{d,y}(\tau)$.
Then, by Prop.~\ref{prop_I_sigma_coincide}, $\sigma$ and $\tau$ differ by a permutation such that each $d$-interval remains in the same affine-linear region of the thermomajorization curve.
If $i\mapsto\frac{y_i}{d_i}$ is injective, this means that there have to exist pairwise distinct $i,j,k\in\{1,\ldots,n\}$ such that both the intervals corresponding to $d_j,d_k$ ``fit inside'' $d_i$.
Therefore $d_i\geq d_j+d_k$ meaning $d$ cannot be well structured. 
\end{proof}

\noindent Note that the proof of Thm.~\ref{thm_necessary_degen} actually shows that $d_i\geq d_j+d_k$ for some distinct $i,j,k$, which is a property stronger than the lack of well-structuredness.
We want to stress that, while condition (i) of Thm.~\ref{thm_necessary_degen} ensures degeneracy of $M_d(y)$ (Lemma~\ref{lem_y_deg}), lack of well-structuredness of $d$ is not sufficient for $M_d(y)$ to be degenerate.
This phenomenon in easy to understand in the graphical representation:
Even if it holds that $d_i\geq d_j+d_k$ for some distinct $i,j,k$---and hence $d$ is not well-structured---it might happen that there is no permutation of the intervals which achieves the degeneracy.
An example of this is given in Sec.~\ref{sec_nondegen_ex}.

\bigskip
Let us conclude this section by having a look at the operator lift. 
More precisely, due to $M_d(y)=\operatorname{conv}\{Ay:A\in\mathrm{ext}(s_d(n))\}$ \cite[Ch.~14, Obs.~C.2.(iii)]{MarshallOlkin}, Minkowski's 
theorem \cite[Thm.~5.10]{Brondsted83} shows that given any extreme point $z$ of $M_d(y)$ there exists 
an extreme point $A$ of $s_d(n)$ such that $z=Ay$.
Now the obvious question
is whether given some extreme point of $M_d(y)$ there is an easy way to recover one (or every) process which maps the initial state to the point in question.
While given any initial and any final state there already exists an algorithm to construct a Gibbs-stochastic matrix mapping the former to the latter \cite{Shiraishi20},
it turns out that if the final state is an extreme point then this procedure simplifies considerably:

\begin{Definition}\label{def_A_sigma_tau}
Given $d\in\mathbb R_{++}^n$ and permutations $\sigma,\tau\in S_n$ there exists, for all $j=1,\ldots,n-1$, a unique $\alpha_j\in\{1,\ldots,n\}$ such that
$
\sum_{i=1}^jd_{\sigma(i)}\in(\sum_{i=1}^{\alpha_j-1}d_{\tau(i)},\sum_{i=1}^{\alpha_j}d_{\tau(i)}]
$.
Also set $\alpha_0:=1$ and $\alpha_n:=n$.
Based on this define a matrix $A_{\sigma\tau}\in\mathbb R_+^{n\times n}$ via
\begin{equation}\label{eq_def_A_sigma_tau}
(A_{\sigma\tau})_{\sigma(j)\tau(k)}:=\begin{cases}
(\sum_{i= 1 }^{\alpha_{ j-1 }  }d_{\tau(i)}-\sum_{i=1}^{j-1}d_{\sigma(i)}  )\cdot d_{\tau(\alpha_{j-1})}^{-1}  & \text{if } k=\alpha_{j-1}<\alpha_j\\
1  &\text{if } \alpha_{j-1}<k<\alpha_{j}  \\
(\sum_{i=1}^jd_{\sigma(i)}-\sum_{i= 1 }^{\alpha_{ j }-1  }d_{\tau(i)}  )\cdot d_{\tau(\alpha_j)}^{-1}  & \text{if } \alpha_{j-1}<\alpha_j=k\\
\frac{d_{\sigma(j)}}{d_{\tau(\alpha_j)}}&\text{if }\alpha_{j-1}=\alpha_j=k\\
0&\text{else}
\end{cases}
\end{equation}
for all $j,k=1,\ldots,n$.
\end{Definition}
This object has already appeared in the literature as ``$\beta$-permutation'' \cite{Alhambra19} and it
coincides with the concept of a ``biplanar extremal transportation matrix'' \cite{Mazurek19} (up to the isomorphism $X\mapsto X{\rm diag}(d)$ from Sec.~\ref{sec_transp_pol}).
The latter name, rightly, suggests the matrix $A_{\sigma\tau}$ for all $y\in\mathbb R^n$, $d\in\mathbb R_{++}^n$, and all $\sigma,\tau\in S_n$ is an extreme point of the Gibbs-stochastic matrices \cite[Thm.~1 ff.]{Mazurek19}.
Yet---due to the lower bound $(n-1)!n^{n-2}$ on the number of extreme points of $s_d(n)$ from Sec.~\ref{sec_transp_pol}---for $n\geq 4$ there must exist extreme points of $s_d(n)$ which are not of the form $A_{\sigma\tau}$ (i.e.~which are not a $\beta$-permutation)\footnote{
Actually, the lower bound in question shows that for large $n$ ``almost no'' extreme point of $s_d(n)$ is a \mbox{$\beta$-permutation} as $(n!)^2/((n-1)!n^{n-2})\to 0$ as $n\to\infty$.
}
\cite[Sec.~IV.B]{Mazurek19}.

Moreover, and more importantly, if $\tau$ is chosen such that $\frac{y_{\tau(1)}}{d_{\tau(1)}}\geq\ldots\geq\frac{y_{\tau(n)}}{d_{\tau(n)}}$, then $A_{\sigma\tau}$ maps the initial state $y$ to the extreme point $E_{d,y}(\sigma)$, and if $y/d$ is non-degenerate, then $A_{\sigma\tau}$ is the \textit{unique} Gibbs-stochastic matrix which maps $y$ to $E_{d,y}(\sigma)$,
cf.~\cite{Alhambra19}, \cite[Lemma~3]{Mazurek19}.
%
Not only does this yield a simple way to reverse-engineer a process which generates an extreme point in question, it also constitutes an alternative way
to evaluate the extreme point map $E_{d,y}$ from Def.~\ref{def_th_E}.


\begin{Remark}
Given a permutation $\sigma$, a matrix $A_{\sigma\tau}$ (stored in a sparse matrix format) can be constructed algorithmically in at most $\mathcal O(n\log n)$ steps using Def.~\ref{def_A_sigma_tau} as the limiting step is to find an appropriate permutation $\tau$. Any algorithm which computes a process matrix $A$ for an arbitrary state transfer, including the one given in~\cite{Shiraishi20}, must have worst time complexity at least $\Omega(n^2)$, since $A$ is dense in general. Hence
the structure of the $A_{\sigma\tau}$ 
leads to an improved runtime
for the special case where the final state is extremal.
\end{Remark}

%

Now our main contribution to this concept reads as follows.
While Definition~\ref{def_A_sigma_tau} (which matches the definition given by~\cite{Alhambra19}) as well as Mazurek's construction for biplanar extremal transportation matrices appear rather convoluted, we will present an incredibly simple construction of this matrix in Sec.~\ref{sec_extr_dstoch} down below.
All one has to do there is to compare the sets $\{\sum_{i=1}^j d_{\sigma(i)}:j=1,\ldots,n\}$ and $\{\sum_{i=1}^j d_{\tau(i)}:j=1,\ldots,n\}$
which, en passant, reaffirms the observation made by Alhambra et al.~that Def.~\ref{def_A_sigma_tau} is independent of the initial state $y$.
Note that these ideas are closely related to the calculation of extreme points given in Sec.~\ref{sec_Mdy} and to the index sets $I_k^\sigma$ defined in~\eqref{eq:intersection-index} which are the main concept in the proof of Prop.~\ref{prop_I_sigma_coincide}.

\section{Detailed Examples}\label{sec_ex}

The objects introduced in Sec.~\ref{sec_degen} can be computed explicitly and they have simple graphical interpretations, e.g., via thermomajorization curves.
The following examples show this in detail.

\subsection{Extreme Point Map} \label{sec_Mdy}

Definition~\ref{def_th_E} contains a simple algorithm to evaluate the extreme point map $E_{d,y}(\sigma)$:
Given some permutation $\sigma\in S_n$ find the value of the thermomajorization curve at $d_{\sigma(1)},d_{\sigma(1)}+d_{\sigma(2)},\ldots,\sum_{i=1}^{n-1}d_{\sigma(i)}$, take the difference of consecutive values, and arrange them into a vector which is ordered according to $\sigma$.
Let us go through a detailed example.

Let $y=(4,0,1)^\top$ and $d=(4,2,1)^\top$. One verifies
$\mathrm{th}_{d,y}(c)=\min\{c,5\}$ for all $c\in[0,7]$ by direct computation, cf.~Figure~\ref{fig_thermocurve_ex_1} below.
Now let $\sigma\in S_3$ be the permutation $\sigma(1)=2$,
$\sigma(2)=3$, and $\sigma(3)=1$; in two-line notation this reads
$\sigma={\scriptsize\begin{pmatrix}1&2&3\\ 2 & 3 &1  \end{pmatrix}}$, henceforth $\sigma=(2\ 3\ 1)$ for short. Our goal 
is to compute the extreme point $E_{d,y}(\sigma)$ of $M_d(y)$ which 
``corresponds'' to $\sigma$. 
We will use the second formulation provided in Definition~\ref{def_th_E}.
First
\begin{align*}
(E_{d,y}(\sigma))_2=(E_{d,y}(\sigma))_{\sigma(1)}&=\mathrm{th}_{d,y}
( d_{\sigma(1)} )-\mathrm{th}_{d,y}(0)=\mathrm{th}_{d,y}( d_2)=\min\{2,5\}=2\,,
\end{align*}
followed by
$(E_{d,y}(\sigma))_3=(E_{d,y}(\sigma))_{\sigma(2)}=\mathrm{th}_{d,y}
( d_{\sigma(1)}+ d_{\sigma(2)} )-\mathrm{th}_{d,y}( d_{\sigma(1)})
=\min\{3,5\}-\min\{2,5\}=1$
and
analogously for
$
(E_{d,y}(\sigma))_1
$.
Thus $E_{d,y}(\sigma)=(2,2,1)^\top$.
This procedure can be nicely visualized,
cf.~Figure~\ref{fig_thermocurve_ex_1}.

\begin{figure}[!ht]
\centering
\includegraphics[width=0.75\textwidth]{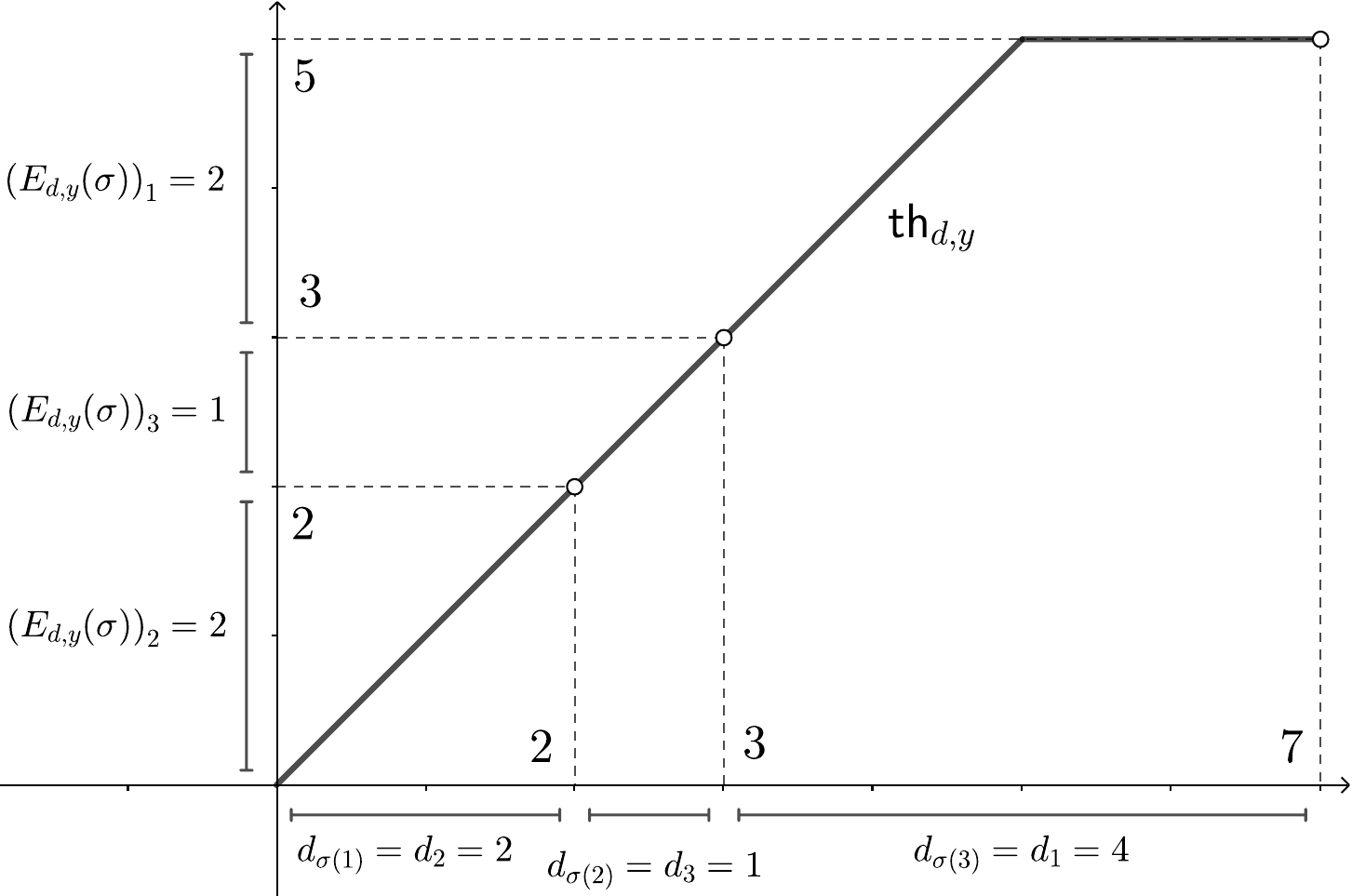}
\caption{
Visualization of the detailed calculation of $E_{d,y}(\sigma)$ for
$\sigma=(2\ 3\ 1)$.
First we extract the value of $\mathrm{th}_{d,y}$ at $d_{\sigma(1)}=d_2=2$ 
which, because we considered the second entry of $d$, becomes the second 
entry of $E_{d,y}(\sigma)$.
Next we add $d_{\sigma(2)}=d_3=1$ to the previous $x$-axis value;
then the third entry of $E_{d,y}(\sigma)$ is the corresponding increment
$\mathrm{th}_{d,y}(3)-\mathrm{th}_{d,y}(2)=1$.
Finally we add $d_{\sigma(3)}=d_1$ to our position on the $x$-axis to arrive at 
${\bf e}^\top d=7$, so the first entry of $E_{d,y}(\sigma)$ is given by the 
final increment $\mathrm{th}_{d,y}(7)-\mathrm{th}_{d,y}(3)=2$.
}
\label{fig_thermocurve_ex_1}
\end{figure}


\noindent The full image of $E_{d,y}$ is computed analogously: one finds 
(cf.~Table~\ref{table1})
$$
\mathrm{ext}(M_d(y))=\Big\{ {\scriptsize \begin{pmatrix} 4\\1\\0 \end{pmatrix},
\begin{pmatrix} 4\\0\\1 \end{pmatrix},\begin{pmatrix} 2\\2\\1 \end{pmatrix},
\begin{pmatrix} 3\\2\\0 \end{pmatrix} } \Big\}\,.
$$

\begin{table}[!ht]
\centering
\begin{tabular}{c|ccc}
$\sigma$ & $\ \mathrm{th}_{d,y}(d_{\sigma(1)})\ $ & $\ \mathrm{th}_{d,y}(d_{\sigma(1)}+d_{\sigma(2)})\ $ & $E_{d,y}(\sigma)$ \\[2mm] \hline\hline\\[-3mm]
$( 1\ 2\ 3)\ $ & $4$ & $\min\{ 6 ,5\}=5$ & $\scriptsize\begin{pmatrix} 4 \\ 5-4 \\ 5-5  \end{pmatrix}=\begin{pmatrix} 4 \\1  \\ 0 \end{pmatrix}$  \\[4mm] \hline\\[-3mm]
$( 3\ 1\ 2)\ $ & $1$ & $\min\{ 5 ,5\}=5$ & $\scriptsize\begin{pmatrix} 5-1 \\ 5-5 \\ 1 \end{pmatrix}=\begin{pmatrix} 4 \\ 0 \\ 1 \end{pmatrix}$  \\[4mm] \hline\\[-3mm]
$(2\ 3\ 1) \ $ & $2$ & $\min\{ 3 ,5\}=3$  &$\scriptsize\begin{pmatrix} 5-3 \\ 2 \\ 3-2 \end{pmatrix}=\begin{pmatrix} 2 \\ 2 \\ 1 \end{pmatrix}$   \\[4mm] \hline\\[-3mm]
$(2\ 1\ 3)\ $ & $2$ & $\min\{ 6 ,5\}=5$  & $\scriptsize\begin{pmatrix}5-2  \\ 2 \\5-5  \end{pmatrix}=\begin{pmatrix} 3 \\ 2 \\ 0 \end{pmatrix}$ \\[4mm] \hline\\[-3mm]
$(1\ 3\ 2)\ $& $4$ & $\min\{ 5 ,5\}=5$  & $\scriptsize\begin{pmatrix} 4 \\5-5  \\ 5-4 \end{pmatrix}=\begin{pmatrix} 4 \\ 0 \\ 1 \end{pmatrix}$  \\[4mm] \hline\\[-3mm]
$( 3\ 2\ 1)\ $ & $1$ & $\min\{ 3 ,5\}=3$ & $\scriptsize\begin{pmatrix} 5-3 \\ 3-1 \\ 1 \end{pmatrix}=\begin{pmatrix} 2 \\ 2 \\ 1 \end{pmatrix}$ \\[4mm] \hline\hline
\end{tabular}
\caption{Image of 
$E_{d,y}$ for $y=(4,0,1)^\top$ and $d=(4,2,1)^\top$ with 
intermediate steps.}\label{table1}
\end{table}

With this in mind let us reformulate the definition of the map $E_{d,y}(\sigma)$ to get an even better understanding:

\begin{Remark}\label{rem_thermo_curves_interpretation}
Given $d\in\mathbb R_{++}^n$ and $\sigma\in S_n$, the permutation $\sigma$ 
tells us how to order the segments of length $d_i$. These we can visualize as 
lying head to tail on the $x$-axis, i.e.~as a tiling of the interval
$[0,{\bf e}^\top d]$. The contact points between the intervals are then
$d_{\sigma(1)}$, $d_{\sigma(1)}+d_{\sigma(2)}$, and so on. Now we evaluate
$\mathrm{th}_{d,y}$ at these points and look at the corresponding increments
\mbox{$(E_{d,y}(\sigma))_{\sigma(j)}=\mathrm{th}_{d,y}( \sum_{i=1}^{j}d_{\sigma(i)} )-\mathrm{th}_{d,y}
( \sum_{i=1}^{j-1}d_{\sigma(i)} )$} for $j=1,\ldots,n$ as visualized
in Figure~\ref{fig_thermocurve_ex_1}.
This construction relates to the previously mentioned notion of tight thermomajorization as these contact points, in turn, are the elbow points of the thermomajorization curve of $E_{d,y}(\sigma)$.

Note that by Definition~\ref{def_th_E}, $(E_{d,y}(\sigma))_{\sigma(j)}$ is the 
increment of $\mathrm{th}_{d,y}$ over a distance of length $d_{\sigma(j)}$. In 
particular for any $j=1,\ldots,n$ the entry $(E_{d,y}(\sigma))_{j}$ corresponds to the 
increment over the interval $d_j$, no matter where it is in our tiling.
Hence two permutations $\sigma,\tau$ give the same extreme point
$E_{d,y}(\sigma)=E_{d,y}(\tau)$ if and only if each interval $d_j$ yields the 
same increment of $\mathrm{th}_{d,y}$ for both permutations.
In the example above this happens because the vector $y/d$ is degenerate.
The following example shows however that this is not necessary.
\end{Remark}

\subsection{Degeneracy} \label{sec_degen_ex} 

Now for a different example: consider $d=(1,2,10)^\top$, $y=(1,4,5)^\top$. 
The key insight is that even though $\frac{y}{d}=(1,2,0.5)^\top$ is non-degenerate, the polytope $M_d(y)$ turns out to be degenerate.
The reason this is allowed to happen is that $d$ is not well structured.
Indeed, as in Sec.~\ref{sec_Mdy} one computes
\begin{equation}\label{eq:th_d_y_ex_2}
\mathrm{th}_{d,y}(c)=\begin{cases}
2c&c\in[0,2]\\
c+2&c\in[2,3]\\
0.5c+3.5&c\in[3,13]
\end{cases}
\end{equation}
for all $c\in[0,13]$ and hence one finds
$$
E_{d,y}(3\ 1\ 2)={\begin{pmatrix}
 \mathrm{th}_{d,y}( 11 )-\mathrm{th}_{d,y}( 10 ) \\ \mathrm{th}_{d,y}( 13 )-\mathrm{th}_{d,y}( 11 ) \\ \mathrm{th}_{d,y}( 10 )
\end{pmatrix}}={\begin{pmatrix}
 9-8.5 \\ 10-9 \\ 8.5
\end{pmatrix}}={\begin{pmatrix}
0.5 \\ 1 \\ 8.5
\end{pmatrix}}
$$
and
$$
E_{d,y}(3\ 2\ 1)=
{\begin{pmatrix}
 \mathrm{th}_{d,y}( 13 )- \mathrm{th}_{d,y}( 12 )  \\  \mathrm{th}_{d,y}( 12 ) - \mathrm{th}_{d,y}( 10 )\\  \mathrm{th}_{d,y}( 10 )
\end{pmatrix}}={\begin{pmatrix}
10-9.5\\ 9.5-8.5 \\ 8.5
\end{pmatrix}}={\begin{pmatrix}
0.5 \\ 1 \\ 8.5
\end{pmatrix}}\,.
$$
Despite the two permutations differing,
their image under $E_{d,y}$ coincides. This comes from the fact that
$\mathrm{th}_{d,y}(10)+\mathrm{th}_{d,y}(13)=\mathrm{th}_{d,y}(11)+\mathrm{th}_{d,y}(12)$ because $\mathrm{th}_{d,y}|_{[3,13]}$ is affine linear, cf.~\eqref{eq:th_d_y_ex_2}. 

\subsection{Non-degeneracy} \label{sec_nondegen_ex} 

The following example shows that even if $d$ fails to be well structured, this does not guarantee that $M_d(y)$ is degenerate.
Indeed, choosing $d=(4,2,1)^\top$, $y=\frac15(3,1,1)^\top$ one computes
$$
\mathrm{ext}(M_d(y))=\frac1{20}\Big\{{\scriptsize \begin{pmatrix}
12\\4\\4
\end{pmatrix},\begin{pmatrix}
13\\4\\3
\end{pmatrix},\begin{pmatrix}
13\\5\\2
\end{pmatrix},\begin{pmatrix}
11\\7\\2
\end{pmatrix},\begin{pmatrix}
10\\7\\3
\end{pmatrix},\begin{pmatrix}
10\\6\\4
\end{pmatrix} }\Big\}
$$
(recall Sec.~\ref{sec_Mdy} for how to evaluate $E_{d,y}$).
In particular \mbox{$|\mathrm{ext}(M_d(y))|=6=3!$} although $d$ is not well structured ($1+2<4$).
The reason for this is fact that when partitioning the interval $[0,7]$ into subintervals of length $(1,4,2)$ (which is the re-ordering of $d$ such that it matches $\frac{y}{d}$ being non-increasing) there is no way to permute these subintervals such that the two small intervals are contained in the big one.

\subsection{Index Sets $I_k^\sigma$} \label{sec_intersections}

\begin{figure}[!ht]
\centering
\includegraphics[width=0.85\textwidth]{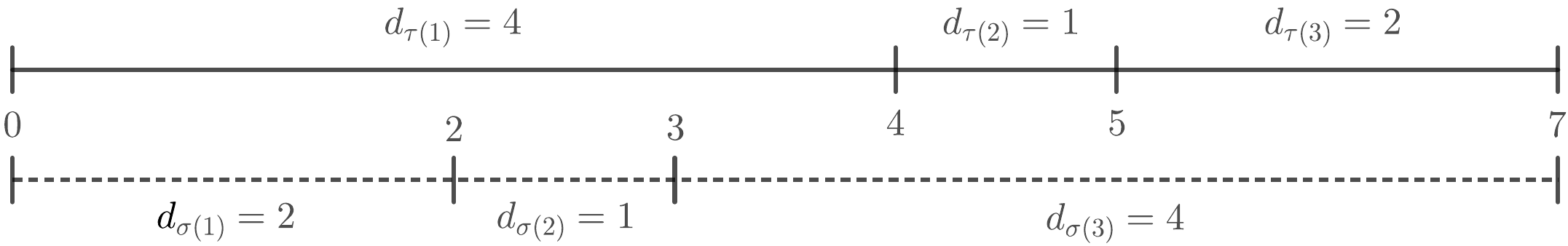}
\caption{
Comparison of the sets $\{\sum_{i=1}^j d_{\tau(i)}:j=1,\ldots,n\}$ (top) and $\{\sum_{i=1}^j d_{\sigma(i)}:j=1,\ldots,n\}$ (bottom) as subsets of the real number line.
}
\label{fig_sorted_intervals_ex_2}
\end{figure}

In~\eqref{eq:intersection-index} we defined the index sets $I_k^\sigma$.
Here we want to see how to easily compute them.
Using the same example as in Sec.~\ref{sec_Mdy}, that is, $y=(4,0,1)^\top$ and $d=(4,2,1)^\top$, we again have the thermomajorization curve in Figure~\ref{fig_thermocurve_ex_1}. 
This curve changes slope only once (i.e.~$m=1$) and it does so at the input $5$;
thus $\Delta_0=0$, $\Delta_1=5$, and $\Delta_2={\bf e}^\top d=7$.
Let us specify the sets $I_\sigma^1,I_\sigma^2$ for the permutation $\sigma=(2\ 3\ 1)$ from Figure~\ref{fig_sorted_intervals_ex_2}.

\noindent One finds $I_1^\sigma=\{\sigma(1),\sigma(2),\sigma(3)\}=\{2,3,1\}$ because all three subintervals intersect the interval $(\Delta_0,\Delta_1)=(0,5)$ corresponding to the first affine linear segment.
Moreover $I_2^\sigma=\{\sigma(3)\}=\{1\}$ because $(\Delta_1,\Delta_2)=(5,7)$ only intersects the third subinterval (w.r.t.~$\sigma$).

\subsection{Extremal $d$-Stochastic Matrices}\label{sec_extr_dstoch}

We already saw that for $y=(4,0,1)^\top$ and $d=(4,2,1)^\top$, the extreme point corresponding to $\sigma=(2\ 3\ 1)$ is $E_{d,y}(\sigma)=(2,2,1)^\top$.
Finding a Gibbs-stochastic matrix which maps $y$ to this extreme point via
Def.~\ref{def_A_sigma_tau} amounts to specifying a permutation which
orders $\frac{y}{d}$ non-increasingly: e.g., choose $\tau=(1\ 3\ 2)$ because then $\underline{\tau}\frac{y}{d}=(\frac{4}{4},\frac{1}{1},\frac{0}{2})^\top=(1,1,0)^\top$.
Now all the information needed for the definition in \eqref{eq_def_A_sigma_tau} is contained in
Figure~\ref{fig_sorted_intervals_ex_2}.

This figure lets us easily build $\underline{\sigma}A_{\sigma\tau}
\underline{\tau}^{-1}$ because the non-zero entries of this matrix correspond to how much of the interval
$(\sum_{i=1}^{j-1}d_{\tau(i)},\sum_{i=1}^{j} d_{\tau(i)}]$ is overlapped by a given 
element of the partition $\{d_{\sigma(i)}:i=1,\ldots,n\}$
(by slight abuse of notation we identify $d_{\sigma(i)}$ with the interval of corresponding length).
For example $(0,d_{\sigma(1)}]$ overlaps with $d_{\tau(1)}$ (covering $\frac24=\,$half of it) but not with $d_{\tau(2)}$, $d_{\tau(3)}$. This means that the first row of $\underline{\sigma}A_{\sigma\tau}
\underline{\tau}^{-1}$ is given by $(\frac12,0,0)^\top$.
Similarly, the second row reads $(\frac14,0,0)^\top$.
On the other hand $d_{\sigma(3)}$ in Fig.~\ref{fig_sorted_intervals_ex_2} overlaps with all three sections $d_{\tau(1)}$, $d_{\tau(2)}$, $d_{\tau(3)}$, and the corresponding overlap ratios are given by $\frac14$, $\frac11$, and $\frac22$. 
Hence the third row of $\underline{\sigma}A_{\sigma\tau}
\underline{\tau}^{-1}$ 
is given by $(\frac14,1,1)^\top$
which altogether yields
$$
\underline{\sigma}A_{\sigma\tau}\underline{\tau}^{-1}=\begin{pmatrix}
\frac12&0&0\\
\frac14&0&0\\
\frac14&1&1
\end{pmatrix}\quad\Leftrightarrow\quad 
A_{\sigma\tau}=\underline{\sigma}^{-1}\begin{pmatrix}
\frac12&0&0\\
\frac14&0&0\\
\frac14&1&1
\end{pmatrix}\underline{\tau}=\begin{pmatrix}
\frac14&1&1\\
\frac12&0&0\\
\frac14&0&0
\end{pmatrix}\,.
$$
One readily verifies that $A_{\sigma\tau}\in s_d(3)$ maps $y$ to $E_{d,y}(\sigma)=(2,2,1)^\top$.
Another observation to make here is that $(A_{\sigma\tau})_{ij}$ is always given by what portion of $d_j=d_{\tau(\tau^{-1}(j))}$ is covered by $d_i=d_{\sigma(\sigma^{-1}(i))}$. In the above example, $(A_{\sigma\tau})_{21}=\frac12$ because half of $d_1=d_{\tau(1)}$ is covered by $d_2=d_{\sigma(1)}$ in Figure~\ref{fig_sorted_intervals_ex_2}.

And there is more to uncover here: on the one hand the $\tau$ we chose is not the only permutation which orders $\frac{y}{d}$ non-increasingly in this example,
and on the other hand we saw in Sec.~\ref{sec_Mdy} that there exists a permutation $\sigma'\in S_3$ other than the $\sigma$ we chose which is mapped to $(2,2,1)^\top$ under $E_{d,y}$.
Thus we can apply the above procedure to $2\cdot 2=4$ combinations of permutations $(\sigma,\tau)$ which all yield extremal Gibbs-stochastic matrices mapping $y$ to $(2,2,1)^\top$. These are given in Table~\ref{table2} below.
\begin{table}[!ht]\label{table_ex_1_continued}
\begin{tabular}{ccc}
$\sigma\downarrow$ / $\tau\rightarrow$ & $(1\ 3\ 2)$ & $(3\ 1\ 2)$ \\[1mm] \hline \\[-3mm]
$( 2\ 3\ 1)$ &$\scriptsize\begin{pmatrix} \frac12&0&0\\
\frac14&0&0\\
\frac14&1&1 \end{pmatrix}$  & $\scriptsize\begin{pmatrix}1&\frac14&0\\
0&\frac14&0\\
0&\frac12&1 \end{pmatrix}$ \\[5mm] \hline \\[-3mm]
$(3\ 2\ 1)$ &$\scriptsize\begin{pmatrix}  \frac14&0&0\\
\frac12&0&0\\
\frac14&1&1 \end{pmatrix}$  &$\scriptsize\begin{pmatrix}1&0&0\\
0&\frac12&0\\
0&\frac12&1\end{pmatrix}$  
\end{tabular}\, 
\begin{tabular}{ccc}
$\sigma\downarrow$ / $\tau\rightarrow$ & $( 1\ 3\ 2)$ & $( 3\ 1\ 2)$ \\[1mm] \hline \\[-3mm]
$( 2\ 3\ 1)$ &$\scriptsize\begin{pmatrix}\frac14  &1  & 1 \\ \frac12 & 0 & 0 \\\frac14  &0  &0  \end{pmatrix}$  & $\scriptsize\begin{pmatrix} \frac12 & 1 & 0 \\ \frac14 & 0 & 1 \\ \frac14 & 0 & 0   \end{pmatrix}$ \\[5mm] \hline \\[-3mm]
$( 3\ 2\ 1)$ &$\scriptsize\begin{pmatrix} \frac14  &1  & 1 \\ \frac12 & 0 & 0 \\\frac14  &0  &0  \end{pmatrix}$  &$\scriptsize\begin{pmatrix} \frac12 & 1& 0 \\ \frac12 & 0 & 0 \\ 0 & 0 & 1 \end{pmatrix}$  
\end{tabular}
\caption{Different combinations of $\sigma,\tau\in S_3$ and the corresponding matrix $\underline{\sigma}A_{\sigma\tau}
\underline{\tau}^{-1}$ (left), as well as $A_{\sigma\tau}$ (right).
}\label{table2}
\end{table}

\section{Conclusions}\label{sec_concl_outl}
In this work, building upon \cite{Lostaglio15_2,Korzekwa17,Lostaglio18,Mazurek19,Alhambra19,Oliveira22} we further explored thermomajorization in the quasi-classical realm as well as the rich geometry of the associated polytope. The former notion comes from the resource theory approach to quantum thermodynamics, and in particular the corresponding set of allowed operations, known as thermal operations.

Inspired by transportation theory the core notions of this work were ``stable'' and ``well-structured'' Gibbs vectors which are simple conditions on the spectrum of a Gibbs state. 
We found that these concepts relate to and give rise to {conceptional insights and} unexpected results regarding system properties and state transfers.
On the one hand, quasi-classical cyclic state-transfers w.r.t.~thermomajorization are impossible if and only if the Gibbs-vector is stable (which is the generic case).
{Put differently, within the model of (quasi-classical) thermal operations performing cyclic state transfers in general comes with a non-zero work cost.}
On the other hand, we uncovered two connections to the notion of sub{space}s in equilibrium:
{1. Thermal operations can bring any sub{space} in equilibrium out of equilibrium without having to expend work. This generalizes the intuitive fact that a system in a Gibbs state can be brought out of equilibrium by coupling it to a non-equilibrium system. Note that for the latter scenario---while any out-of-equilibrium system is necessarily a resource---our result shows that this is the only price one has to pay, i.e.~once the systems are coupled there is some process on the composite system which takes the original system out of the Gibbs state and which can be implemented at no work cost.
2. The existence of sub{space}s in equilibrium is} reflected in the geometry of the thermomajorization polytope.
Indeed, assuming well-structuredness of the Gibbs state---which is equivalent to the system's temperature exceeding a critical value---the existence of a sub{space} in equilibrium corresponds to the polytope having degenerate corners.

While we explored the case of quasi-classical states in all detail, as usual in quantum thermodynamics the general case is a lot more difficult. 
Most notably, the number of extreme points of the thermal operations as well as
the number of extreme points of
the future thermal cone is uncountable for non-classical initial states
\cite{Lostaglio15,vomEnde22thermal}.
Hence one loses access to tools from the theory of convex polytopes.
One of the few things that pertain to general systems are the notions of well-structured and stable Gibbs states (Def.~\ref{def_well_ord_stable});
investigating whether these notions encode any properties of general thermal cones (such as, e.g., the impossibility of cyclic processes) could be an interesting topic for future research.
%
%
%
%
%
%

%
%
\vspace{6pt} 

\authorcontributions{Conceptualization, F.v.E. and E.M.; methodology, F.v.E. and E.M.; software, F.v.E. and E.M.; investigation, F.v.E. and E.M.; writing---original draft preparation, F.v.E. and E.M.; writing---review and editing, F.v.E. and E.M.; visualization, F.v.E. and E.M.; supervision, F.v.E. All authors have read and agreed to the published version of the manuscript.}

\dataavailability{No new data were created or analyzed in this study. Data sharing is not applicable to this article.} 



\acknowledgments{
{We are grateful to thank Gunther Dirr and Thomas Schulte-Herbr\"uggen
for valuable and constructive comments during the preparation of this manuscript.
Moreover, we would like to thank the the anonymous referees for their helpful comments.}
F.v.E.~has been supported by the Einstein Foundation (Einstein Research Unit on Quantum Devices) and the MATH+ Cluster of Excellence.
E.M.~is part of the Bavarian excellence network \textsc{enb}
via the International PhD Programme of Excellence
\textit{Exploring Quantum Matter} (\textsc{exqm}), as well as the \textit{Munich Quantum Valley} of the Bavarian
State Government with funds from Hightech Agenda \textit{Bayern Plus}.
}

\conflictsofinterest{The authors declare no conflict of interest.} 



\abbreviations{Abbreviations}{
The following abbreviations are used in this manuscript:\\

\noindent 
\begin{tabular}{@{}ll}
w.l.o.g. & without loss of generality
\end{tabular}
}

\appendixtitles{no} 
\appendixstart
\appendix
\section[\appendixname~\thesection]{Basic Properties of Concave Functions}\label{app_concave_lemma}
We start with the following simple observation:
Given a compact interval $I\subseteq\mathbb R$, a concave function $f:I\to\mathbb R$, and $r,s,t\in I$ with $r<s<t$ one finds
$$
f\Big( \frac{s-r}{t-r}t  +\frac{t-s}{t-r}r \Big)\geq  \frac{s-r}{t-r}f(t)  +\frac{t-s}{t-r}f(r)
$$
which in turn is equivalent to
\begin{equation}\label{eq:concave_slope}
\frac{f(t)-f(s)}{t-s}\leq\frac{f(s)-f(r)}{s-r}\,.
\end{equation}
The following now is a direct consequence of this:

\begin{Lemma}\label{lemma_th_ordered}
Let $d\in\mathbb R_{++}^n$, $y\in\mathbb R^n$, $\sigma\in S_n$. One has $\mathrm{th}_{d,y}(\sum_{i=1}^jd_{\sigma(i)})=\sum_{i=1}^j y_{\sigma(i)}$ for all $j=1,\ldots,n-1$ if and only if $\frac{y_{\sigma(1)}}{d_{\sigma(1)}}\geq\ldots\geq\frac{y_{\sigma(n)}}{d_{\sigma(n)}}$.
\end{Lemma}
\begin{proof}
``$\Leftarrow$'': Shown in \cite[Lemma 15 (iii)]{vomEnde22}. ``$\Rightarrow$'': 
Because $\mathrm{th}_{d,y}$ is concave \cite[Lemma 15 (i)]{vomEnde22} we may insert $r=\sum_{i=1}^{j-1}d_{\sigma(i)}$, $s=\sum_{i=1}^{j}d_{\sigma(i)}$, $t=\sum_{i=1}^{j+1}d_{\sigma(i)}$ for $j=2,\ldots,n-1$ in \eqref{eq:concave_slope}.
This yields the desired inequality.
\end{proof}

Moreover, the following basic property of concave functions will be essential for our analysis of thermomajorization curves:

\begin{Lemma}\label{lemma_concave_connecting_line}
Let $I\subset\mathbb R$ be a compact interval, that is, $I=[a,b]$ for some $a,b\in\mathbb R$, $a<b$. Given $f:I\to\mathbb R$ concave or convex the following statements are equivalent.
\begin{itemize}
\item[(i)] $f$ is affine linear, i.e.~for all $x\in[a,b]$
$$
f(x)=\frac{x-a}{b-a}f(b)+\frac{b-x}{b-a}f(a)\,.
$$
\item[(ii)] For all $y_1,y_2\in(a,b)$ which satisfy $y_1+y_2=a+b$ one has
$
f(y_1)+f(y_2)=f(a)+f(b)
$.
\item[(iii)] There exist $y_1,y_2\in(a,b)$ such that $y_1+y_2=a+b$ and
$
f(y_1)+f(y_2)=f(a)+f(b)
$.
\item[(iv)] There exists $y\in(a,b)$ such that $(y,f(y))$ lies on the line which connects $(a,f(a))$ and $(b,f(b))$. More precisely,
$$
f(y)=\frac{y-a}{b-a}f(b)+\frac{b-y}{b-a}f(a)\,.
$$
\end{itemize}
\end{Lemma}
\begin{proof}
``(i) $\Rightarrow$ (ii)'': Straightforward calculation.
``(ii) $\Rightarrow$ (iii)'': Trivial.
``(iii) $\Rightarrow$ (iv)'': Assume w.l.o.g.~that $f$ is concave---the convex case is done analogously. First note that concavity implies
\begin{equation}\label{eq:concavity_basic_property}
f(x)=f\Big(\frac{x-a}{b-a}b+\frac{b-x}{b-a}a\Big)\geq \frac{x-a}{b-a}f(b)+\frac{b-x}{b-a}f(a)
\end{equation}
for all $x\in[a,b]$. This leads to
\begin{equation}\label{eq:concavity_basic_property_2}
f(a)+f(b)=f(y_1)+f(y_2)\geq f(y_1)+\frac{y_2-a}{b-a}f(b)+\frac{b-y_2}{b-a}f(a)
\end{equation}
and thus
\begin{align*}
\frac{y_1-a}{b-a}f(b)+\frac{b-y_1}{b-a}f(a)&\overset{\eqref{eq:concavity_basic_property}}\leq f(y_1)
\\&\overset{\eqref{eq:concavity_basic_property_2}}\leq \Big(1-\frac{y_2-a}{b-a}\Big)f(b)+\Big(1-\frac{b-y_2}{b-a}\Big)f(a) \\
&\overset{\hphantom{\eqref{eq:concavity_basic_property_2}}}= \frac{b-a-y_2+a}{b-a}f(b)+\frac{b-a-b+y_2}{b-a}f(a)\\
&\overset{\hphantom{\eqref{eq:concavity_basic_property_2}}}=\frac{y_1-a}{b-a}f(b)+\frac{b-y_1}{b-a}f(a)  \,.
\end{align*}
which is what we had to show. In the last step we used $y_1=a+b-y_2$ (because $y_1+y_2=a+b$ by assumption).

``(iv) $\Rightarrow$ (i)'': Again assume w.l.o.g.~that $f$ is concave---the convex case is done similarly. Let $x\in(a,y)$ be given. Using the assumption on $f(y)$ as well as concavity we compute
\begin{align*}
\frac{y-a}{b-a}f(b)+\frac{b-y}{b-a}f(a)
&=f(y)=f\Big( \frac{y-x}{b-x}b+\frac{b-y}{b-x}x \Big)\geq  \frac{y-x}{b-x}f(b)+ \frac{b-y}{b-x}f(x) 
\end{align*}
and thus
\begin{align*}
f(x)&\leq \frac{b-x}{b-y}\Big( \frac{y-a}{b-a}f(b)+\frac{b-y}{b-a}f(a)-\frac{y-x}{b-x}f(b)  \Big)\\
&= \frac{(b-y)(x-a)}{(b-y)(b-a)}f(b)+\frac{b-x}{b-a}f(a)=\frac{x-a}{b-a}f(b)+\frac{b-x}{b-a}f(a)
\end{align*}
by means of a straightforward computation. The converse inequality comes from~\eqref{eq:concavity_basic_property}, hence equality holds. The case $x\in(y,b)$ is done analogously which concludes the proof.
\end{proof}

%
%

\begin{adjustwidth}{-\extralength}{0cm}

\reftitle{References}


\bibliography{control21vJan20}

\PublishersNote{}
\end{adjustwidth}
\end{document}